\documentclass[onecolumn,notitlepage,showpacs,superscriptaddress,,amsmath,amssymb,aps,pre]{revtex4-1}

\usepackage{graphics}
\usepackage{graphicx}
\usepackage{dcolumn}
\usepackage{bm}

\newcommand*{\beq}{\begin{equation}}
\newcommand*{\eeq}{\end{equation}}
\newcommand{\Teleia}{\quad .}
\newcommand{\Comma}{\quad ,}

\newcommand{\p}{\partial}

\renewcommand{\l}{\langle}
\renewcommand{\r}{\rangle}
\newcommand{\f}{\frac}
\newcommand{\s}{\sum}

\newcommand{\rro}{\right)}
\newcommand{\lro}{\left( }
\newcommand{\lsq}{\left[}
\newcommand{\rsq}{\right]}
\newcommand{\rcu}{\right\}}
\newcommand{\lcu}{\left\{}
\newcommand{\be}{\begin{equation}}
\newcommand{\ee}{\end{equation}}
\newcommand{\bi}{\begin{itemize}}
\newcommand{\ei}{\end{itemize}}
\newcommand{\ben}{\begin{enumerate}}
\newcommand{\een}{\end{enumerate}}
\newcommand{\esp}{ESPResSo }
\newcommand{\rmin}{r_{\textrm{min}}}
\newcommand{\rcut}{r_{\textrm{cut}}}
\newcommand{\umin}{u_{\textrm{min}}}
\newcommand{\usigma}{u_{\sigma}}
\newcommand{\umod}{u_{mm}^{\textrm{mod}}}

\newcommand{\jas}{J. Aerosol Sci.}

\newcommand{\bea}{\begin{eqnarray}}
\newcommand{\eea}{\end{eqnarray}}

\newcommand{\commentout}[1]{{}}

\newcommand{\hv}{\harvarditem}

\begin{document}
%

\title{Langevin agglomeration of nanoparticles interacting via a central potential}

\author{Lorenzo Isella}
\altaffiliation{Present address: ISI Foundation, Turin 10133, Italy}
\email{lorenzo.isella@isi.it}
\affiliation{European Commission, Joint Research Centre,
I-21027 Ispra (VA), Italy}

\author{Yannis Drossinos}
\thanks{Corresponding author}
\email{ioannis.drossinos@jrc.ec.europa.eu}
\affiliation{European Commission, Joint Research Centre,
I-21027 Ispra (VA), Italy}
\affiliation{School of Mechanical \& Systems Engineering,
Newcastle University, Newcastle upon Tyne NE1 7RU, United Kingdom}

\date{\today}

\begin{abstract}
Nanoparticle agglomeration in a quiescent fluid is simulated by
solving the Langevin equations of motion of a set of
interacting monomers in the continuum regime. Monomers interact via a
radial, rapidly decaying intermonomer potential.
The morphology of generated clusters is analyzed through their
fractal dimension $d_f$ and the cluster coordination number.
The time evolution of the cluster fractal dimension is linked
to the dynamics of two populations,
small ($k \le 15$) and large ($k>15$) clusters.
At early times monomer-cluster agglomeration is the dominant
agglomeration mechanism ($d_f = 2.25$), whereas at late times
cluster-cluster agglomeration dominates ($d_f = 1.56$).
Clusters are found to be compact (mean coordination number
$\sim 5$), tubular, and elongated.
The local, compact structure of the aggregates is attributed to the
isotropy of the interaction potential, which allows rearrangement of
bonded monomers, whereas the large-scale
tubular structure is attributed to its relatively short attractive range.
The cluster translational diffusion coefficient is determined to be
inversely proportional to the cluster mass and the (per-unit-mass)
friction coefficient of an isolated monomer, a consequence of the neglect of
monomer shielding in a cluster. Clusters generated by unshielded
Langevin equations are referred to as \textit{ideal clusters}
because the surface area accessible to the
underlying fluid is found to be the sum of the accessible surface
areas of the isolated monomers. Similarly, ideal clusters do
not have, on average, a preferential orientation. The decrease of
the numbers of clusters with time and a few collision kernel
elements are evaluated and compared to analytical expressions.
\end{abstract}
%


\maketitle

\section{Introduction}
\label{introduction}

Nanoparticle aggregates are of significant importance in
technological and industrial processes such as, for example,
combustion, filtration, and gas-phase particle synthesis. In addition,
colloidal aggregates play an important role in, for example,
the pharmaceutical industry, food processing, paintings, and polymers.
The fractal nature of these aggregates
has profound implications on their
transport \cite{filippov_drag} and thermal \cite{filippov_thermal} properties.
Fractal aggregates arise from the agglomeration of smaller units,
herein taken to be spherical and referred to as monomers, that do
not coalesce but rather retain their identity in the resulting
aggregate. In a quiescent fluid the main mechanism driving
agglomeration is diffusion: accordingly individual monomers may be
modelled as interacting Brownian particles whose motion and dynamics
is described by a set of Langevin equations. Langevin simulations
have been used to study aggregate formation
\cite{mountain}, aggregate collisional properties
\cite{pratsinis_kernel}, the limits of validity of the Smoluchowski
equation \cite{pratsinis_ld}, and aggregate films
\cite{friedlander_deposition}.

In this study we investigate nanoparticle agglomeration
and the diffusive motion and growth of the resulting aggregates
by relying solely on the Langevin equations of motion
of a set of interacting monomers in three dimensions.
The monomer-monomer interaction potential is taken to be
either a model potential, composed of a repulsive and a
short-ranged attractive part, or a Lennard-Jones intermolecular potential integrated
over the monomer volumes.
Both potentials are spherically symmetric and rapidly decaying.
Spherical symmetry implies that intermonomer forces are central:
no angular force limits reorientation
of an attached monomer. We shall argue that these two features
have profound implications on the small and large-scale
structure of the generated aggregates.
Unlike the works mentioned above,
no assumptions are made about the structure (e.g., by specifying the
cluster fractal dimension) or the mobility of the
aggregates.
Our approach is reminiscent of other applications of Langevin simulations in
dilute colloidal suspensions to study diffusion-induced agglomeration and the
structures it gives rise to \cite{videcoq}. Stochastic agglomeration (particle motion
and deposition) has also been discussed in
terms of Brownian dynamics, see, for example, Refs. \cite{Hutter, Biswas}.

An inherent difficulty of the adopted mesoscopic description
of nanoparticle agglomeration, a description whereby the
effect of collisions of fluid molecules with larger solid nanoparticles
is modelled by a random force, is the choice of the stochastic properties of the random force.
These properties, in particular the noise strength,
are usually specified by invoking the Fluctuation Dissipation Theorem (FDT).
We will use the FDT to relate the amplitude of
the fluctuations of the random force acting on a monomer to
its friction coefficient. In applying the FDT, however,
monomer shielding will be neglected in that the friction
coefficient of a monomer in an aggregate will be
taken to be independent of its state of aggregation.
This approximation of the hydrodynamic
forces acting on a monomer has been referred to as
the free draining approximation~\cite{TangentialForces}.
 The monomer friction and diffusion coefficients
are related to the monomer surface area accessible (or exposed)
to surrounding fluid molecules~\cite{ActiveSurface}. The accessible surface area
is the fraction
of the geometric surface area that is active in momentum and energy
transfer from the underlying fluid to the monomer: it,
thus, determines the monomer and aggregate transport properties.
We will argue that the accessible surface area of a generated $k$-monomer cluster
(also referred to as a $k$-mer),
is $k$ times the accessible surface area of an isolated monomer, i.e.,
the accessible surface area of a sphere.
We shall, thus, refer to the clusters generated by unshielded Langevin equations
as \textit{ideal clusters} with respect to their dynamic properties.
Monomer shielding may, equivalently, be described in terms of the cluster
shielding factor defined as the ratio of the average (per-unit-mass)
friction coefficient of a monomer
in an aggregate to the (per-unit-mass) friction coefficient of an isolated
monomer.

In a non-ideal cluster the monomer accessible surface area decreases due
to shielding
leading to a decrease of the cluster friction coefficient, and
a consequent increase of its diffusion coefficient.
\citet{filippov_thermal} presented
an expression for the shielding factor of a cluster by
comparing the total heat transfer to an aggregate to the product of the
number of monomers in the aggregate times the heat transfer to
an isolated monomer (under the same conditions).
Monomer shielding may be understood by noting that the fluid concentration boundary
layers of neighbouring monomers in a cluster overlap non-additively, thereby the
fluid molecule-monomer collision rate decreases.
In a Langevin description of cluster diffusion the change
of the monomer accessible surface, and hence the change of the cluster
shielding factor, renders the strength of the
stochastic thermal noise a time-dependent function of the local arrangement
of each monomer. We do not attempt such a modification
of the noise strength, adopting the form of the FDT that leads to
unshielded Langevin equations.

Since we solve numerically the Langevin equations for interacting monomers
rather than the aggregate equations of motion, information
on the dynamics of aggregate formation has to be inferred
from the simulation output. We obtain this datum, together with the
detailed structure of each aggregate, from the record of the collisions
it underwent and its eventual restructuring,
using techniques borrowed from graph theory \cite{book_algorithms}.

The paper is organized as follows: Section \ref{model} provides the
theoretical framework for the Langevin simulations.
Emphasis is placed on the description and justification of the
monomer-monomer potentials used in the numerical experiments.
Section \ref{sec:simulations} describes the numerical method, and
it introduces the quantities monitored to investigate the statics and
dynamics of the generated aggregates. Section \ref{sec:Statics}
presents the static properties of the aggregates, namely their
fractal dimension, their morphology (specifically, the cluster coordination number),
and some indications of cluster restructuring. Section \ref{sec:Dynamics} summarizes
the cluster dynamic properties, in particular, their translational
diffusion coefficient, their mean Euler angles, the decay of the total number of clusters with time, and
the determination of a limited number of
agglomeration kernel elements. The final remarks in Sec.~\ref{conclusions}
summarize the main results and conclude the paper.

\section{Model description}
\label{model}

\subsection{Monomer Langevin equations of motion}
\label{sec:lang-equat-mesosc}

We investigate the non-equilibrium dynamics of interacting clusters in
the continuum regime (fluid mean free path smaller than the monomer
diameter) by
solving the Langevin  equations of motion of a dilute system of $N$ interacting monomers
in a quiescent fluid
in three dimensions. We use the word cluster with the same meaning as the term
aggregate in Ref. \cite{konstandopoulos}, i.e., a set of physically bound
spherules (monomers). The $i$-th monomer obeys the Langevin equation
\begin{equation}
\label{eq:Langevin}
m_1\ddot{\bf r}_i={\bf F}_i-\beta_1 m_1\dot{\bf r}_i+
{\bf W}_i(t) \quad , \quad i = 1, \ldots N \Comma
\end{equation}
where $m_1$ in the monomer mass, ${\bf r}_i$ its position in
three-dimensional space, ${\bf F}_i$ an external force,
$\beta_1$ the monomer friction coefficient per unit mass,
and ${\bf W}$ a noise term that
models the effect of collisions between the monomer
and the molecules of the surrounding quiescent fluid.
We consider that each monomer feels a Stokes drag (continuum regime),
the friction coefficient per unit mass being $\beta_{1} = 3 \pi \mu_f \sigma/m_1$ where $\mu_f$
is the fluid viscosity, and $\sigma$ the monomer diameter. Henceforth
all friction coefficients will be defined per unit mass.
The monomer friction coefficient may also be expressed
as the inverse monomer relaxation
time $\beta_1 = \tau_1^{-1}$ where $\tau_1 = \rho_p \sigma^2/(18 \mu_f)$ with
$\rho_p$ the monomer material density. As argued in
the Introduction, the use of Stokes drag implies that
the monomer surface area accessible to fluid molecules equals the
accessible surface area of a monomer irrespective of its state of aggregation.
The implications
of this approximations are explored in Sec.~\ref{sec:clust-diff-coeff}.
The noise is assumed to be Gaussian and white (delta-correlated in
time)
\begin{equation}
\label{eq:gaussian_noise}
\langle  W^{k}_i(t) \rangle=0 \Comma \quad {\rm{and}} \quad \langle
W^{k}_i(t) W_{j}^{l}(t')\rangle = \Gamma \delta_{ij}
\delta_{k l}\delta(t-t') \Comma
\end{equation}
where angular brackets $\langle \ldots \rangle$ denote an ensemble average
over realizations of the random force, subscripts denote monomer number ($i, j = 1, \ldots N$), and
superscripts Cartesian coordinates ($k,l = x,y,z$). The noise strength
$\Gamma$ is determined from the Fluctuation Dissipation Theorem
applied to each monomer: it evaluates to $\Gamma=2\beta_1 m_1k_BT$
with $k_B$ Boltzmann's constant and $T$ the system temperature~\cite{risken_book}.

The force acting on the $i$-th monomer arises from its interaction with all the other
monomers. It is considered to be conservative
\begin{equation}
\label{eq:potential_pairwise}
{\bf F}_i=- \mbox{\boldmath$\nabla$}_{{\bf r}_i} U_i \Comma
\end{equation}
where $U_i$ is the total intermonomer potential
the $i$-th monomer feels. We assume that it derives from a pair-wise additive,
two-body, radial potential $u_{ij}(r_{ij})$
\begin{equation}
\label{eq:pairwise_pot_explicit}
U_i =  \s_{j\neq i}^N u_{ij}(r_{ij}) \Comma
\end{equation}
where the radial distance
is $r_{ij}=|{\bf r}_i-{\bf r}_j|$.

Equations~(\ref{eq:Langevin}) may be cast in dimensionless form, a form
more convenient for their numerical solution. By choosing
length $l^*$, time $\tau^*$, and  mass $m^*$ characteristic
scales we introduce the dimensionless variables, denoted by tilde,
\begin{equation}
\label{eq:dimensionless-scales}
r\equiv l^*\tilde r \Comma \quad t \equiv \tau^*\tilde t \Comma \quad
m_1\equiv m^*\tilde m_1 \Teleia
\end{equation}
These characteristic scales fix the characteristic system temperature $T^*$ to
\beq
k_BT^*\equiv\f{m^*(l^*)^2}{(\tau^*)^2} \Teleia
\label{eq:CharacteristicTemperature}
\eeq

Equations~(\ref{eq:Langevin}) and~(\ref{eq:gaussian_noise}) in dimensionless
and component-wise form become
\begin{subequations}
\label{eq:LangevinDimensionless}
\begin{equation}
\label{eq:Langevin-dimensionless-componentwise}
\f{d^2\tilde r_i^l}{d\tilde t^2} =
-\f{1}{\tilde m_1}\f{\p\tilde U}{\p \tilde r_i^l}
-\tilde\beta_1\f{d\tilde r_i^l}{d\tilde t}
+\f{1}{\tilde m_1}\tilde W_i^l(\tilde t) \quad ; \quad
\ l=x,y,z \ \quad ; \quad  i = 1, \ldots N
\Comma
\end{equation}
\begin{equation}
\langle \tilde W_i^l(\tilde t)\rangle=0 \quad , \quad {\rm{and}} \quad
\langle \tilde W_i^l(\tilde t) \tilde W_j^k(\tilde t')
\rangle= 2 \ \tilde \beta_1 \tilde m_1 \tilde T \delta_{ij}
\delta_{kl} \delta(\tilde t-\tilde t')
\Comma
\label{eq:NoiseDimensionless}
\end{equation}
\end{subequations}
where we introduced the dimensionless variables
\begin{equation}
\label{eq:derived-dimensionless-quantities}
\tilde\beta_1\equiv \beta_1\tau^* \Comma \quad
\tilde T=\f{T}{T^*} \Comma \quad
\tilde U\equiv \f{U}{k_BT^*} \Comma \quad
\tilde W_i^l\equiv\f{(\tau^*)^2}{m^*l^*}W_i^l \Teleia
\end{equation}
Equations~(\ref{eq:LangevinDimensionless}) show that
three independent dimensionless variables ($\tilde m_1$,
$\tilde \beta_1$, $\tilde T$) determine the dynamics of the
system.
A natural, but not unique, choice of units
is $m^*=m_1$, $\tau^*=\tau_1$, and $l^*=\sigma$, where $\sigma$
is a characteristic length scale of the intermonomer potential
(to be taken to be the monomer diameter).  This
choice leads to $\tilde\beta_1=\tilde m_1=1$ in
Eq.~(\ref{eq:Langevin-dimensionless-componentwise}), a form of the Langevin
equations we will use in our numerical simulations. All simulations were
performed at $\tilde{T} = 0.5$.

The characteristic scales may be evaluated for a typical case
of agglomeration of combustion-generated nanoparticles. A typical
soot monomer of material density $\rho_p=1.3$g/cm$^3$ and
characteristic size $l^{\star} = \sigma=20$nm has a characteristic
temperature $T^*$
\begin{equation}
\label{eq:T_star}
T^{\star} = \f{18^2\pi\mu_f^2\sigma}{6k_B\rho_p} \simeq 650{\rm K}
\Comma
\end{equation}
when suspended in air at $300$K of dynamic viscosity
$\mu_f=1.85\cdot 10^{-4}$g/(cm$\cdot$sec).
Therefore, $\tilde T = 0.5$ corresponds to approximately $T \simeq 300$K.
The Stokes monomer relaxation time, the characteristic time scale,
is $\tau_1 \sim 1.6 \times 10^{-9}$sec. Heat, mass, and momentum transfer
between particles and the carrier gas depend on the Knudsen number
$\textrm{Kn} = 2 \lambda_g/\sigma$, where $\lambda_g$ is the carrier gas
mean free path: for $\textrm{Kn} \ll 1$ these transfer processes occur
in the continuum regime. The air mean free path at atmospheric pressure
and $293$K is $\lambda_{\textrm{air}} = 66$nm. Hence, our simulations are
appropriate either for
aerosol agglomeration at high pressures ($\lambda_g \sim p_g^{-1}$, with $p_g$
the carrier gas pressure) or for agglomeration of non-charged colloids in liquids.

\subsection{Monomer-monomer interaction potential}
\label{explain_potential}

The monomer-monomer potential is chosen to mimic
interaction of hard-core monomers sticking upon collision.
As such it will be taken to be rapidly decaying. The effect of the
range of repulsive interactions in two-dimensional colloidal aggregation
was extensively discussed in Ref.~\cite{TwoDRangeRepulsive}, whereas
Ref.~\cite{videcoq} studied the effect of the attractive range.
The early studies \cite{mountain, meakin_cluster_models}
considered perfect sticking of two monomers when their relative
distance fell below the monomer diameter. In these works the authors did not use
an interaction potential, but they examined the system
frequently during its time evolution to identify agglomeration events
(monomer-cluster or cluster-cluster) by calculating the relative distances of
all monomers. After, e.g., two clusters had touched, the
relative distances of all the monomers in the resulting cluster were
``frozen'', and the cluster was allowed to diffuse with a diffusion
coefficient that had to be prescribed \textit{a priori}.

Herein, aggregate formation arises from monomer collisions
that bind them through their interaction.
We used two spherically symmetric
intermonomer potentials: a model potential $u_{mm}^{\textrm{mod}}$
and a potential $u_{mm}$ that arises from the integration
of the intermolecular potential over
the volume of two macroscopic bodies (e.g., two monomers).

\begin{center}
\begin{figure*}
\includegraphics[width=\columnwidth]{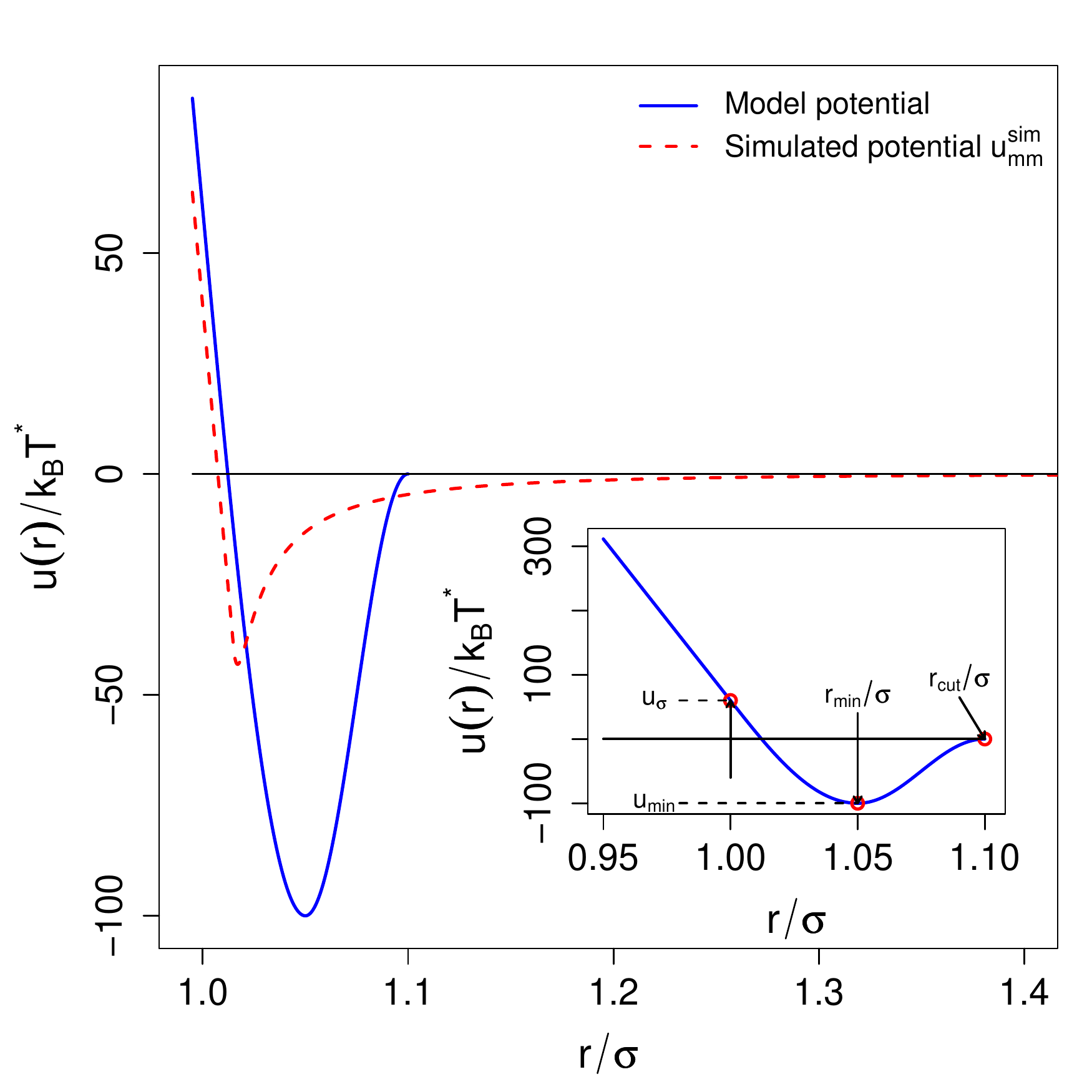}
\caption{Model and simulated monomer-monomer interaction potentials.}
\label{plot_potential}
\end{figure*}
\end{center}

The  model potential $u_{mm}^{\textrm{mod}}$ is short ranged,
and it has a
deep and narrow attractive well
to model monomer binding without break-up. Furthermore, it tends smoothly to zero at
$r= r_{\rm cut}$ where $r_{\rm cut}$ is a cut-off distance
such that $r_{\rm cut}-\sigma\ll\sigma$. This avoids the introduction
of the so-called impulsive forces in the system \cite{md_book}.
It is attractive on a length scale much smaller than
the monomer diameter $\sigma$.
We chose the following analytical expression
\begin{subequations}
\label{eq:potential_well}
\begin{eqnarray}
\umod(r) = \frac{\pi}{2} \, \Big ( \frac{\usigma - \umin}{\rmin - \sigma} \Big )
\left ( r - \sigma \right ) + \usigma & \quad \text{if} \quad & 0<r \le \sigma \Comma \\
\umod(r) = \usigma - \left ( \usigma - \umin \right ) \, \cos \Big [ \frac{\pi}{2} \,
\left ( \frac{\rmin -r}{\rmin - \sigma} \right ) \Big ]
& \quad \text{if} \quad & \sigma<r \le r_{\rm min} \Comma \\
\umod(r) = \umin \cos^2 \Big [ \frac{\pi}{2} \, \left ( \frac{r -\rmin}{\rcut - \rmin} \right ) \Big ]
& \quad \text{if} \quad & r_{\rm min}<r \le r_{\rm cut} \ , \\
u_{mm}^{\textrm{mod}} (r) = 0 & & \quad \rm{elsewhere} \Comma
\end{eqnarray}
\end{subequations}
where $\usigma = \umod (\sigma)$, $\rmin$ is the location of the potential minimum,
and $\umin = \umod (\rmin)$. The model potential depends on five parameters
($\rmin$, $\sigma$, $r_{\textrm{cut}}$, $\usigma$, and $\umin$)
chosen in our numerical simulations as follows. The potential minimum
is located at $r_{\rm min}=1.05\sigma$, where it
evaluates to $u_{\rm min}=-100k_BT$.
At monomer separation $\sigma$ the potential evaluates to
$u (\sigma) =60k_BT$ with a steep gradient. For monomer separations in
the range $(0,\sigma]$ the potential is extrapolated linearly
with the slope it has at $r=\sigma$.
Hence, for separations smaller than $\sigma$, the monomers feel a
strong constant repulsive force equal to the force at $r=\sigma$.
The cut-off distance is $r_{\textrm{cut}}=1.1\sigma$.
The model potential and its various parameters are shown in
the inset of Fig.~\ref{plot_potential}.

The interaction potential
may be obtained from the integration of the
intermolecular interactions over the nanoparticle volumes
as in Ref.~\cite{yannis_potential}. We assume pairwise additivity
of the intermolecular potential, continuous medium, and  constant material
properties. Elastic flattening of the monomer is neglected.
The intermolecular potential is
taken to be the Lennard-Jones potential
\begin{equation}
\label{eq:LJ}
u_{LJ}(r)=4\epsilon\lsq\lro\f{\sigma_{LJ}}{r}\rro^{12}-\lro\f{\sigma_{LJ}}{r}\rro^6  \rsq
\Comma
\end{equation}
where $\epsilon$ is the depth of the attractive potential,
the maximum attractive energy between two molecules, and $\sigma_{LJ}$
the distance at which the potential evaluates to zero,
the distance of closest approach of two molecules which collide
with zero initial relative kinetic energy. The first
term expresses (approximately) the repulsive part (Born repulsion),
the second the attractive (van der Waals attraction).
Integration of Eq.~(\ref{eq:LJ}) over two
equal-sized spheres of diameter $\sigma$ yields, see, for example,
Refs.~\cite{yannis_potential, ruckenstein}, an attractive part
\begin{subequations}
\label{eq:Umm}
\beq
\label{eq:UmmAttractive}
u_{mm}^{\textrm{vdW}} (r) = - \frac{A}{6}
\ \lsq \ln \lro \f{r^2-\sigma^2}{r^2} \rro
 + \f{\sigma^2}{2(r^2-\sigma^2)} + \f{\sigma^2}{2r^2} \rsq \Comma
\eeq
and a repulsive part
\beq
\label{eq:UmmRepulsive}
\begin{split}
u_{mm}^{\textrm{rep}} (r) = {} & \frac{A \sigma_{LJ}^{6}}{2520r}
\lcu \sigma^2 \lsq \f{1}{2(r-\sigma)^7} + \f{1}{2(r+\sigma)^7}+\f{1}{r^7} \rsq \right. \\
& \left. - \f{\sigma}{3} \lsq \f{1}{(r-\sigma)^6}- \f{1}{(r+\sigma)^6} \rsq
-  \f{1}{15} \lsq \f{2}{r^5}-\f{1}{(r-\sigma)^5}-\f{1}{(r+\sigma)^5} \rsq \rcu \Comma
\end{split}
\eeq
to obtain
\beq
\label{eq:UmmSum}
u_{mm} (r) = u_{mm}^{\textrm{rep}} (r) + u_{mm}^{\textrm{vdW}} (r) \Comma
\eeq
\end{subequations}
where $A=4\pi\epsilon\sigma_{LJ}^6n^2$ is the Hamaker constant and
$n$ the molecular number density in the solid. In the
limit $r \gg \sigma$ the potential decays as
$u_{mm} \sim - r^{-6}$, i.e., the expected
attractive van der Waals interaction energy between
two non-charged macroscopic bodies is recovered. The Hamaker constant
of a typical soot nanoparticle, for example $n$-hexane,
was estimated to be $A=2.38\cdot 10^{-19}$J \cite{kasper_hamaker}
with a corresponding $\sigma_{LJ}=0.5949$nm \cite{substance_book}. We used these
values, along with $\sigma = 20$nm, to render the interaction
potential dimensionless.

The repulsive part of $u_{mm}$ diverges for $r\to\sigma$; it
may, thus, cause numerical difficulties in the solution of the
Langevin equations.
We modified the potential
at distances $r_{\rm mat}=1.015\sigma$ (less than the position
of the potential minimum at $r_{\min}\simeq 1.017\sigma$) by
extrapolating it linearly with the same slope it has at
$r_{\rm mat}$ until $r=0.995\sigma$ (where it evaluates to about $60 k_B T^*$).
A similar matching condition was used to determine the coefficient
of the model-potential linear term. We set $u_{mm}$ to zero at smaller monomer-monomer separations.
This modification of the repulsive part
is not expected to affect the dynamics of the system, since it
involves monomer separations that are energetically unfavorable
(and, hence, unlikely).
Therefore, the intermonomer potential we use is
\begin{subequations}
\label{eq:van_der_waals}
\begin{eqnarray}
u_{mm}^{\textrm{sim}} (r) = 0 & \quad \text{if} \quad & 0 \le r \le 0.995 \sigma \Comma \\
u_{mm}^{\textrm{sim}} (r) = \f{\p u_{mm}}{\p r}\bigg{|}_{r_{\rm mat}} (r_{\rm mat}-r)+u_{mm}(r_{\rm mat})
& \quad \text{if} \quad & \quad 0.995 \sigma <r \le r_{\rm mat} \Comma \\
u_{mm}^{\textrm{sim}} (r) = u_{mm}(r) & \quad \text{if} \quad &  \quad r_{\rm mat} < r \Teleia
\end{eqnarray}
\end{subequations}
We truncate the potential at distances
$r \geq 7 \sigma$ where it is negligible with respect to the thermal energy,
$u_{mm} (7\sigma)/(k_BT^*)\sim 10^{-4}$.

Although neither potential diverges at separations $r<\sigma$,
at such distances monomers feel a very strong repulsive force;
monomer separations below $\sigma$ are energetically
unfavorable and their occurrence is extremely unlikely during the system dynamics. This
justifies the identification of $\sigma$ with the hard-core monomer diameter.
Moreover, since the two intermonomer potentials are much deeper
than $k_BT^{\star}$ the
sticking probability upon collision may be considered to be unity.

In the following, we will use only dimensionless quantities;
we will, thus, drop the tilde to simplify notation. Furthermore,
unless specified otherwise, the results presented were
obtained with $u_{mm}^{\textrm{sim}}$, Eqs.~(\ref{eq:van_der_waals}).
Figure~\ref{plot_potential} presents and compares the
two dimensionless radial potentials.

\section{Numerical method}
\label{sec:simulations}

\subsection{Numerical solver}
\label{sec:Solver}

The numerical simulations were performed with the software
package ESPResSo \cite{espresso}, a versatile package for generic Molecular Dynamics
simulations in condensed-matter physics. We used the Molecular Dynamics
program with a Langevin thermostat.
The Langevin thermostat
was construed as formal method to perform Molecular Dynamics simulations
in a constant temperature canonical ensemble, see,
for example, Ref.~\cite{Thermostat}.
It introduces a fictitious viscous force to model the coupling of
the system to a thermal bath according to Eqs.~(\ref{eq:Langevin})
that ensures the system temperature fluctuates about
the bath mean temperature.
The molecular equations of motion
with a Langevin thermostat are formally identical to the Langevin equations
of interacting Brownian particles: the physical scales and their interpretation
differ.


The  ESPResSo numerical solver uses the Verlet algorithm for the
deterministic part, with a numerical error that scales at least like
$O (\delta t^3)$. The combination of the Verlet algorithm with the
solver for the stochastic part (Langevin thermostat) yields an error
estimate of $O (\delta t^3)$ in the monomer positions and of order
$O (\delta t^2)$ in the velocities \cite{NumericsEspresso}. As a
check of the numerical solver we simulated the motion of a single
Brownian particle in a quiescent fluid. The simulations for the
mean-square displacement $\langle r_{1}^2(t) \rangle$, the
mean-square velocity fluctuations $\langle v_1^2 (t) \rangle$, and
the velocity autocorrelation function $\langle {\bf v}_1(t){\bf
v}_1(0) \rangle$ agreed with the analytical
expressions~\cite{risken_book}.

The initial state was created by randomly placing $n_{\infty}(0)V=5000$
monomers in a cubic box of size $L$
with $V = L^3$ the box volume, and $n_{\infty}(0)$ the initial
monomer concentration. The box size was chosen to give the desired initial monomer
concentration according to $L=\lro{5000}/{n_{\infty}}\rro ^{1/3}$.
We chose $n_{\infty}(0)=0.01$ corresponding to $L\simeq 80$.
The initial random placement of monomers could cause
numerical problems if two or more monomers happen to be placed in
almost overlapping positions, thus experiencing immediately a very
strong repulsive force. We used a well-known technique of
MD simulations to avoid these numerical instabilities by ramping up the repulsive force
for $r<1$ to its constant final value  during 800 time steps.
The (dimensionless) simulation time step was chosen to be $\delta t_{\textrm{sim}} = 1.25 \times 10^{-3}$.
After initialization, the system was evolved until
a final time $3000$, when the cluster concentration
had decreased by almost two orders of magnitude.
The results we show were obtained
using the output of $10$ simulations, each one with different initial monomer positions and
zero initial monomer velocities.
Finally, periodic boundary conditions were imposed.

\subsection{Cluster identification}
\label{sec:cluster-calculation}

The identification of clusters is one of the most time-consuming
tasks of post-processing the simulation results. The \esp simulations
return individual monomer positions and velocities. Unlike the previously
mentioned works \cite{mountain, meakin_cluster_models},
agglomeration events are not identified during the time evolution of
the system, but cluster formation is determined \textit{a posteriori}. Sampling
of the  simulation output was performed at time intervals $\delta
t_{\rm sam}=2$ (every $1,600$ simulation time steps).

We resort to an approach based on graph theory.
A cluster is a set of connected, bound, monomers.
As both interaction potentials are very deep once two monomers
collide and bind, they remain bound: no agglomerate break-up was noticed
during our simulations.
Hence, two monomers may be considered bound when their
relative distance is less than a threshold distance $d_{\rm thr}$.
The choice of $d_{\rm thr}$ depends on the position of the minimum of
monomer interaction potential $r_{\rm min}$. For the simulations with
the integrated Lennard-Jones potential we used $d_{\rm thr}=1.04$,
whereas for the model potential we used $d_{\rm thr}=1.06$.
Small variations of $d_{\rm thr}$ about
these values did not affect the identification of clusters.
Our definition of a cluster is reminiscent of the liquid cluster definition
proposed by Stillinger and used in gas-liquid
nucleation studies (see, for example, Ref.~\cite{clusterReguera}).

Computationally, the first step is the calculation of the distance
matrix ${\bf D}$ between all monomers.
For a box with periodic boundary conditions, the distance between two
monomers is the distance between the $i$-th monomer and the nearest
image of the $j$-th monomer \cite{md_book}. For instance, the
(ordered) distance between two monomers along coordinate $l$ is
\begin{equation}
\label{eq:distance_calc}
D_{ij}^{l}= r_i^{l}-r_j^{l}-L\cdot{\rm nint}\lro\f{r_i^{l}-r_j^{l}}{L} \rro
\quad , \quad l=x,y,z \Comma
\end{equation}
where ``nint" is the nearest integer function. The periodicity of the
box imposes a cut-off of $L/2$ on the maximum distance between two monomers along each axis.
The three-dimensional distance
\begin{equation}
\label{eq:distance_3D}
D_{ij} = \Big [ \sum_{l=1}^3 \big ( D_{ij}^{l} \big )^2 \Big ]^{1/2} \Comma
\end{equation}
is always a non-negative quantity. The distance matrix $\bf{D}$
is subsequently used to calculate the adjacency
matrix $\bf{A}$, defined as
\begin{equation}
\label{eq:adjacency_matrix}
A_{ij} =
\begin{cases}
1 & \quad \text{if } \quad D_{ij} \le d_{\rm thr} \Comma \\
0 & \quad \text{otherwise} \Teleia
\end{cases}
\end{equation}

The adjacency matrix is usually introduced in graph theory \cite{book_algorithms}
as a convenient way to represent a graph uniquely. Monomers in a cluster can be formally regarded as
graph vertices (nodes), whereas the bonds due to the interaction potential become
graph edges (links). The problem of cluster identification, given the distance matrix
${\bf D}$, is then re-formulated as the identification of
the connected components of a non-directed graph expressed by the
(symmetric) adjacency matrix ${\bf A}$. They can be determined using a
standard breadth-first search algorithm
\cite{book_algorithms}. Due to its speed and scalability, we resort
to the implementation of its algorithm in the igraph library \cite{igraph}.

\subsection{Cluster radius of gyration}
\label{sec:ClusterRg}

The radius of gyration is a geometric parameter used
to characterize the size of fractal aggregates. It describes the spatial
mass distribution about the aggregate center of mass. As such
it is a static property that depends on the cluster mass distribution, and
not on the diffusive properties of the cluster.
For a cluster composed of $k$ equal-mass monomers
it becomes the root-mean-square distance of the monomers from
the cluster center of mass~\cite{filippov_thermal}
\begin{subequations}
\begin{equation}
\label{eq:r_gyr_definition}
R_g^2 = \frac{1}{k} \, \s_{i=1}^k{ ({\bf r}_i -{\bf R}_{CM})^2} + a_1^2
\Comma
\end{equation}
where the aggregate center of mass is
\beq
{\bf R}_{CM}= \frac{1}{k} \, \s_{i=1}^k{\bf r}_i \Comma
\label{eq:Rcm}
\eeq
\end{subequations}
and $a_1$ a monomer characteristic size. Since we are interested
in the power-law dependence of the radius of gyration
on the number of monomers in a cluster even for small clusters we included $a_1$ in
the definition of the radius of gyration: otherwise Eq.~(\ref{eq:r_gyr_definition})
evaluates to zero for a monomer. The additional term may be taken to
be either the primary particle radius~\cite{filippov_thermal}, $a_1 = \sigma/2$,
or the radius of gyration of a sphere~\cite{sorensen_prefactor},
$a_1=\sqrt{3/5}(\sigma/2)$. The choice of $a_1$ as the monomer radius
of gyration ensures that, if an
average fractal exponent is used, the large $k$ behaviour
in the power-law scaling persists even for smaller clusters. Of course,
the value of $a_1$ (including $a_1=0$) is irrelevant for large
clusters. We used both choices for $a_1$: the results presented
were obtained with $a_1$ the monomer radius of gyration because
this choice gave better agreement of the calculated and numerically
estimated kernel elements, see Sec.~\ref{sec:KernelElementCalculate}.
Nevertheless, the difference between the two choices was small.

The calculation of $R_g$ according to Eq.~(\ref{eq:r_gyr_definition})
must take into consideration the periodicity of the box.
Since all monomer-monomer distances are
known from Eq.~(\ref{eq:distance_calc}), the position of monomers
in the cluster with respect to a randomly selected monomer (say monomer $1$) may be easily calculated;
for instance, the $j$-th monomer position along coordinate $l$
with respect to monomer $1$ is
\begin{equation}
\label{eq:reconstruct_cluster}
 r_j^{l} =r_1^{l} +D_{1j}^{l}  \quad \Teleia
\end{equation}
Given the relative position of all monomers in the cluster, the
position of the aggregate center of mass
may be calculated via Eq.~(\ref{eq:Rcm}), and the radius of gyration
from Eq.~(\ref{eq:r_gyr_definition}).
We adopted a reference system centered on the randomly chosen
monomer in the cluster, i.e., ${\bf  r}_1=(0,0,0)$.
For this reference system $ r_j^{l} = D_{1j}^{l}$.
This procedure is independent of the choice of the selected
monomer, since the radius of gyration  is independent of the cluster
position in the box,
nor does the procedure depend on the sign convention chosen for $D_{1j}^{l}$.

\subsection{Collision kernel}
\label{sec:Kernel}

The package \esp allows addressing each monomer individually
at all times during the simulation, i.e. a permanent label (e.g., color) may
be associated with each monomer. A monomer label
allows the unequivocal identification of the cluster it belong to.
Each cluster becomes an unordered
collection of monomer labels where no monomer label is repeated: this amounts
to the mathematical definition of a set.
Viewing clusters as sets of monomer labels provides a computational method
to investigate cluster collisions even for sampling times considerably
longer than the simulation time step, as long as the aggregates do not break-up.
To be specific, consider two clusters at time $t$.
If during the (sampling) time interval $(t, t+\delta t_{\textrm{sam}}]$ they collide, they will be
part of the same newly-formed cluster at  $t+\delta t_{\textrm{sam}}$.
An equivalent description of the collision is that the two monomer-label sets that
identify the pre-collision clusters become proper subsets of
the monomer-label set that identifies the cluster formed by their collision, and detectable at
$t+\delta t_{\textrm{sam}}$. The number of collisions that occurred during
$(t, t+\delta t_{\textrm{sam}}]$,
and the clusters involved in the collisions, may be recorded simply by comparing different sets.

The collision kernel $K_{ij}$ (rendered dimensionless by scaling it by
$\sigma^3/\tau_1$) between an $i$-mer and $j$-mer during the sampling time interval is
\begin{equation}
\label{eq:kernel_calc}
\f{N_{ij}(t)}{V \delta t_{\textrm{sam}}} = (2-\delta_{ij}) K_{ij} \f{N_i(t)}{V}
\f{N_j(t)-\delta_{ij}}{V}
\cong (2-\delta_{ij}) K_{ij}n_i(t)n_j(t) \Comma
\end{equation}
where $N_{ij}(t)$ is the number of collisions between $i$- and $j$-mers
that took place during the interval $(t, t+\delta t_{\textrm{sam}}]$,
$N_i(t)$ is the number of $i$-mers at time $t$, $n_i(t)=N_i(t)/V$ the cluster concentration,
and $\delta_{ij}$ the Kroenecker delta. The collision kernel $K_{ij}$ is estimated
from Eq.~(\ref{eq:kernel_calc}) by treating it as the fitting parameter
that minimizes (in a least-square sense) the distance between the
number of detected collisions
$N_{ij}(t)/(V \delta t_{\textrm{sam}})$ and $(2-\delta_{ij}) K_{ij}n_i(t)n_j(t)$.

\section{Static cluster properties}
\label{sec:Statics}

Examples of 50-monomer clusters that survive at the end of one of the simulations
are shown in Fig.~\ref{snapshot_end_simulation}. The top, left subfigure is a
snapshot of the system, whereas the other three
are color-coded aggregates. The color code denotes the number of
first neighbours of a monomer. Note the relatively compact and long, tubular
structure of the generated aggregates. These two features,
small-scale compactness and large-scale tubular structure, will be related
to properties of the intermonomer potential in the following sections.

\subsection{Cluster fractal dimension}
\label{sec:calculate_df}

The fractal dimension of the clusters is determined
from the statistical scaling law that governs the power-law
dependence of the cluster radius gyration $R_g$ on cluster size $k$
\begin{equation}
\label{eq:determination_df}
R_g = a k^{1/d_f} \Comma
\end{equation}
where $d_f$ is the average, time-independent fractal dimension of the aggregates,
and $a$ the fractal prefactor, occasionally referred to as
lacunarity~\cite{Rosner95}. The fractal prefactor provides information
on the packing of monomers~\cite{WuFriedlander93}. The cluster radius of
gyration was obtained by averaging over
generated cluster configurations. Specifically, the average radius of
gyration of a $k$-monomer aggregate is taken over all $k$-monomer aggregates
that had been recorded at least $200$ times. This requirement aims at
eliminating outliers. It
is not particularly stringent since for each simulation $1500$ system configurations (snapshots)
were stored, and $10$ simulations. Our results are not sensitive to reasonable choices
of the threshold of minimum number of cluster occurrences: raising the occurrence threshold
to $400$ does not change the calculated fractal dimension.

\begin{figure*}
\includegraphics[width=\columnwidth]{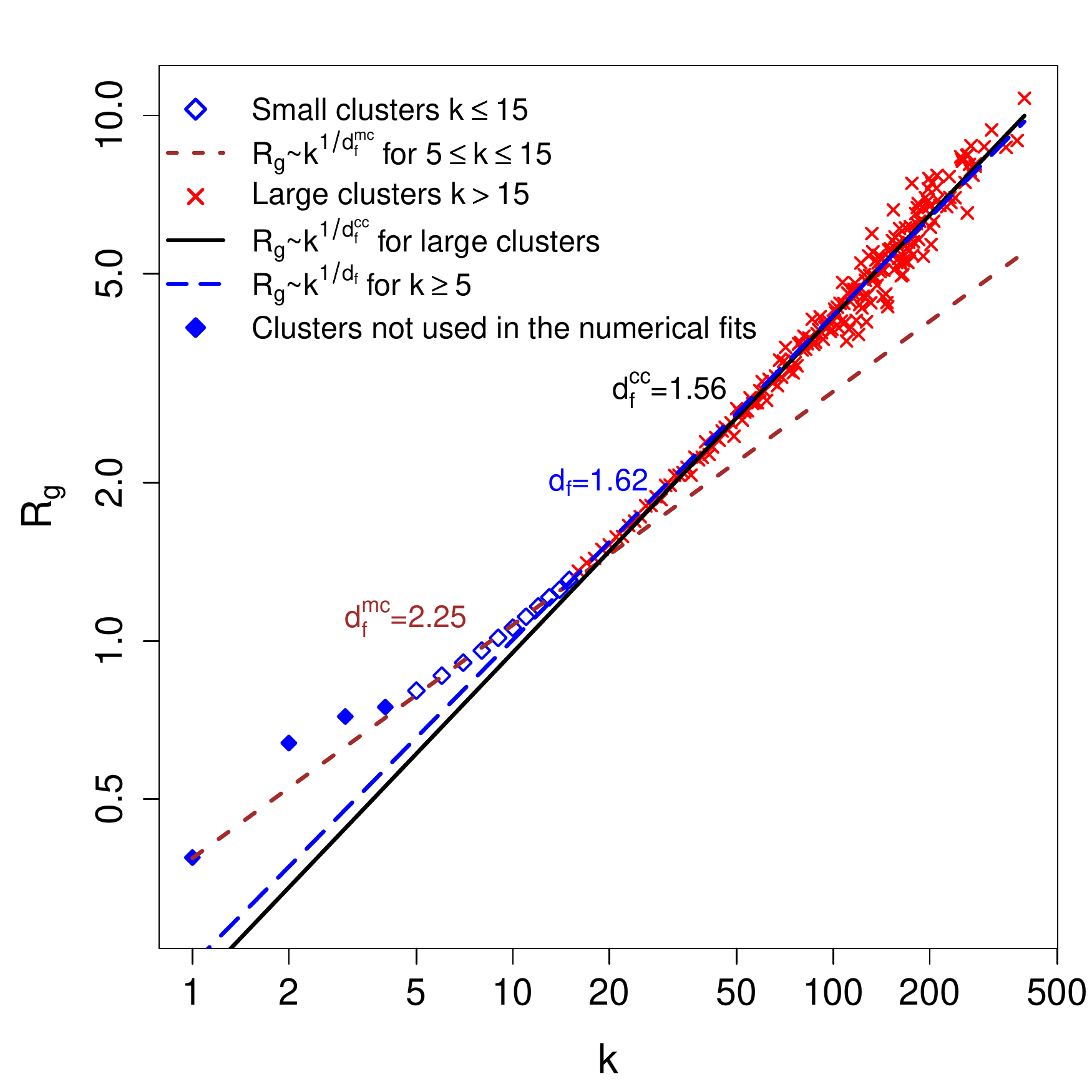}
\caption{Average radius of gyration as a function of cluster
size. Linear fits performed on a double-logarithmic scale
for $k>5$ (long-dashed line), $5\le k\le 15 $ (short-dashed line)
and $k>15$ (solid line) (see, also, Table~\ref{tab:RadiusScaling}).}
\label{two_df}
\end{figure*}

Figure~\ref{two_df} presents the calculated radius of
gyration as a function of aggregate size. The double logarithmic plot
shows that power-law scaling breaks down
for cluster sizes $k<5$. Hence, we set $k=5$ the minimum cluster size in the fits.
All cluster configurations with $k \ge 5$
were fitted to a single line leading to an average fractal exponent $d_f=1.62 \pm 0.02$,
and a prefactor $a=0.242 \pm 0.006$. The prefactor re-expressed
in terms of $k = k_g (2R_g)^{d_f}$
gives $k_g = 3.24$. The calculated prefactor
is on the high side of fractal prefactors reported
in the literature for clusters generated by
computer simulations~\cite{PrefactorAeroSci00}, but closer to experimentally
determined prefactors that
indicate $k_g > 2$ ~\cite{PrefactorAeroSci00,Rosner95}.
Inspection of the figure shows that clusters with $k\le 15$ deviate
significantly from the single-line fit. Simulation results
are fitted better by considering two cluster population (and hence
two different slopes): we identify small ($k\le 15$) and
large ($k>15$) clusters. Refitting the data gives
a fractal dimension $d_f^{mc}=2.25 \pm 0.05$ for small clusters
and $d_f^{cc}=1.56 \pm 0.02$ for large clusters. The corresponding
average prefactors, as well as the fractal
dimensions, are reported in Table~\ref{tab:RadiusScaling}.
As before, the prefactor for the large clusters are on the high side
of literature-reported values for computer-generated clusters, and
closer to experimental values determined by angular light
scattering, suggesting that the cluster generated in this study have
a closely packed structure~\cite{WuFriedlander93}, see also
Fig.~\ref{snapshot_end_simulation}.

The presence of two cluster populations obeying scaling laws with
different exponents may be attributed to different
agglomeration mechanisms. Small clusters
are generated mainly by monomer-cluster agglomeration;
they are more compact and spherical than large clusters.
This agglomeration process is similar to, but different from,
diffusion-limited aggregation, as argued in Sec.~\ref{sec:coordination-number}.
Large clusters are mainly generated by cluster-cluster agglomeration.
Therefore, it is expected, and numerically confirmed, that
$d_f^{mc}>d_f^{cc}$. Furthermore,
the calculated fractal dimension of large clusters is comparable, but lower,
to reported fractal dimensions of cluster-cluster agglomeration
$d_f \sim 1.7 - 1.8$~\cite{mountain, smirnov_review}. Further comments
on the agglomeration process and the associated fractal dimensions
are made in Sec.~\ref{sec:coordination-number}.

\begin{figure*}
\includegraphics[width=\columnwidth]{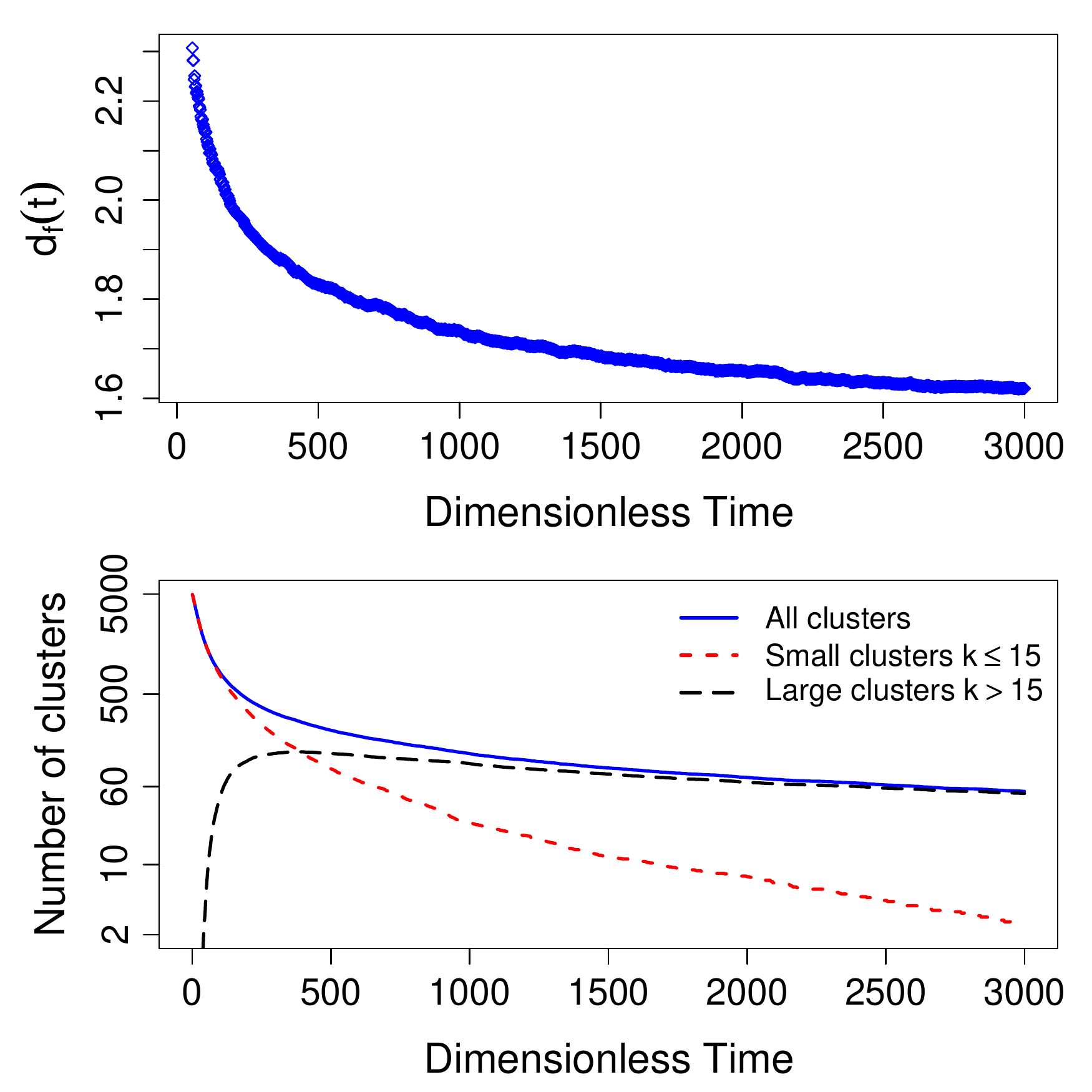}
\caption{Top: Aggregate fractal dimension as a function of
time. Bottom: Total number of clusters (solid), number of small clusters $k \leq 15$
(short dashed), and number of large clusters $k > 15$ (long dashed).}
\label{evolution_df}
\end{figure*}

The existence of two cluster populations that arise from
the predominance of different agglomeration mechanisms as
agglomeration progresses suggests
the calculation of a time-dependent average fractal dimension
$d_f (t)$. We calculated it by considering the instantaneous radius of gyration
of $k$-monomer clusters for each of the ten initial configurations:
only clusters with $k \ge 5$ were included, and no occurrence threshold
was imposed.
As the number of clusters is considerably smaller than those used in
the time-independent calculation we
do not calculate first the instantaneous average radius of gyration and
then fit the data (as done for the calculation presented in Fig.~\ref{two_df}):
instead we fit all the data simultaneously (no averaging).
The fractal dimension as a function of time is shown in Fig.~\ref{evolution_df} (top).
The evolution of the fractal dimension may be linked to the kinetics of
the small and large cluster populations. At early times, the system is almost entirely made
up of small clusters ($k \le 15$), Fig.~\ref{evolution_df} (bottom),
hence $d_f (t) \simeq d_f^{mc}$, whereas at late times
the contribution of large clusters ($k > 15$) to the overall cluster population
is dominant, hence $d_f (t) \to d_f^{cc}$.

These findings are in the spirit of the study by Kostoglou and
Konstandopoulos~\cite{konstandopoulos} who used a distribution of fractal
dimensions to characterize aggregate morphology. They showed that
the mean fractal dimension relaxes to an \textit{a priori} fixed
asymptotic value specified by the dominant agglomeration mechanism.
Our results on the importance of
agglomeration kinetics on the evolution of $d_f (t)$ are in
agreement with earlier works, e.g., Refs.~\cite{mountain, smirnov_review}
who suggest that the aggregate fractal dimension
depends mainly on the dominant agglomeration mechanism.


\subsection{Cluster morphology}
\label{sec:coordination-number}


As a measure of local compactness of the generated
aggregates we calculated the cluster coordination number.
The cluster coordination number, defined as the mean number of first
neighbours of a monomer in a cluster, provides information on the openness of the
aggregates and their compactness, and on the presence of cavities in their structure,
i.e., their porosity. Hence, it is related to the fractal prefactor,
Eq.~(\ref{eq:determination_df}); it characterizes the
small-scale structure of an aggregate, whereas the fractal
dimension characterizes its large-scale structure.
The coordination number varies between 0 and 12 for spherical particles,
reaching its highest value for hexagonal close packed (hcp) or
face centered cubic (fcc) structures at a volume fraction of 0.74 \cite{Hutter}.
The cluster coordination number
is obtained during post-processing of the
simulation results resorting again to the igraph library
\cite{igraph}.

\begin{figure*}
\includegraphics[width=\columnwidth]{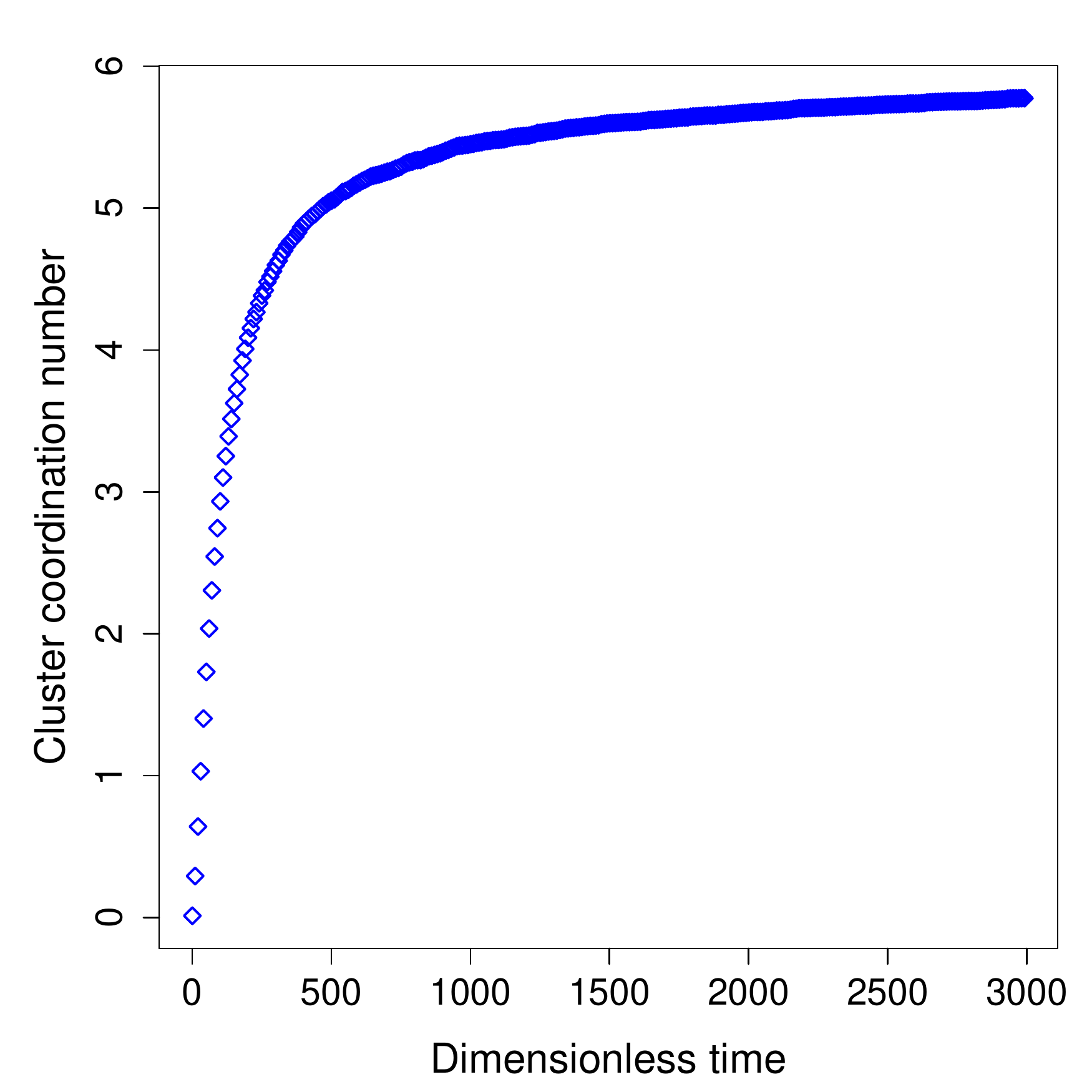}
\caption{Mean cluster coordination number.}
\label{evolution_coord_number}
\end{figure*}

We calculated the mean cluster coordination number
by averaging the coordination number of each cluster
over all clusters at a given time.
It is plotted in Fig.~\ref{evolution_coord_number} as a function of time.
At late times it reaches values higher than five, implying that
the aggregates are relatively compact.

\begin{turnpage}
\begin{figure*}
\includegraphics[width=0.60\columnwidth,height=0.40\columnwidth]{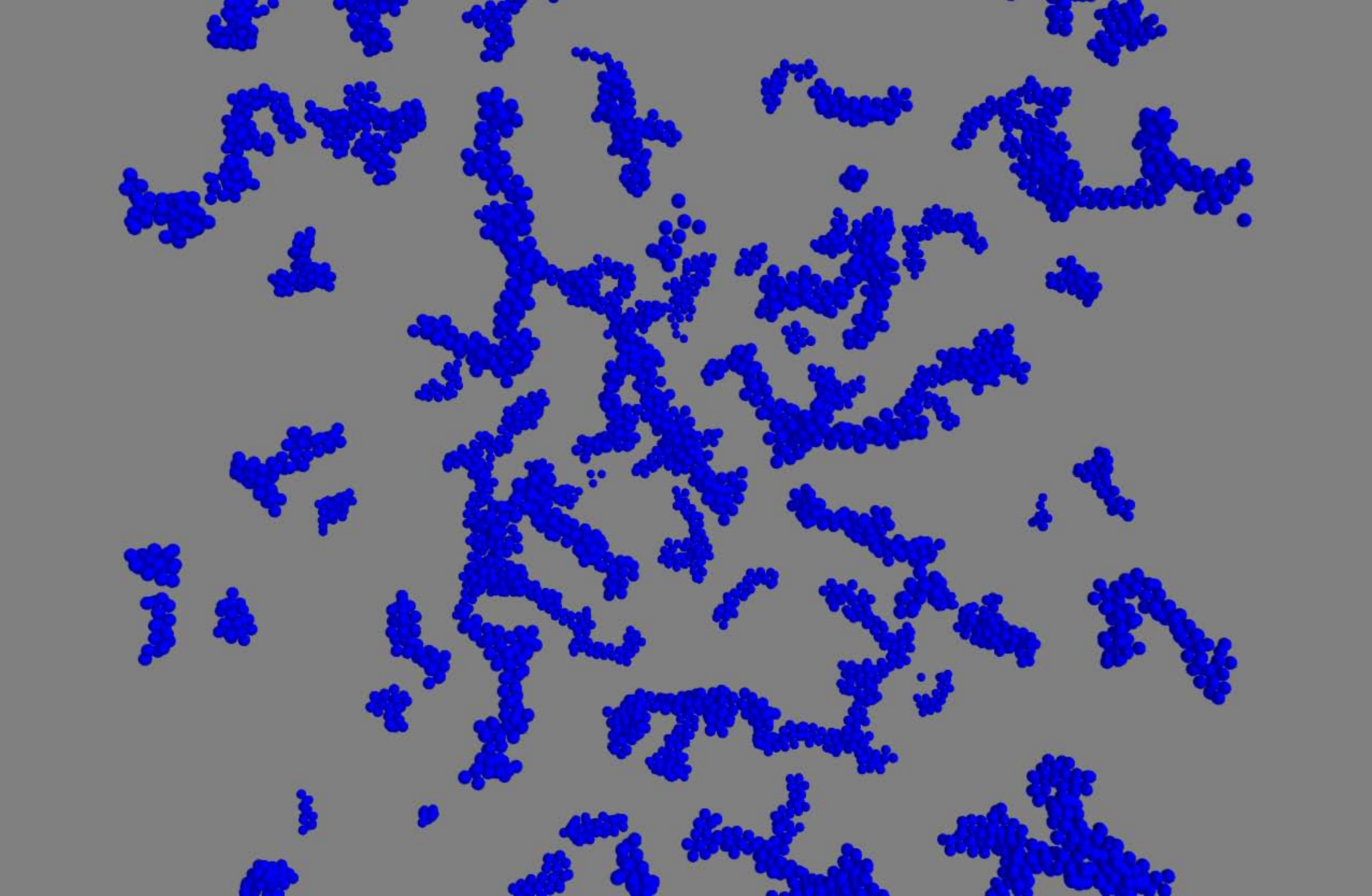}
\includegraphics[width=0.60\columnwidth,height=0.40\columnwidth]{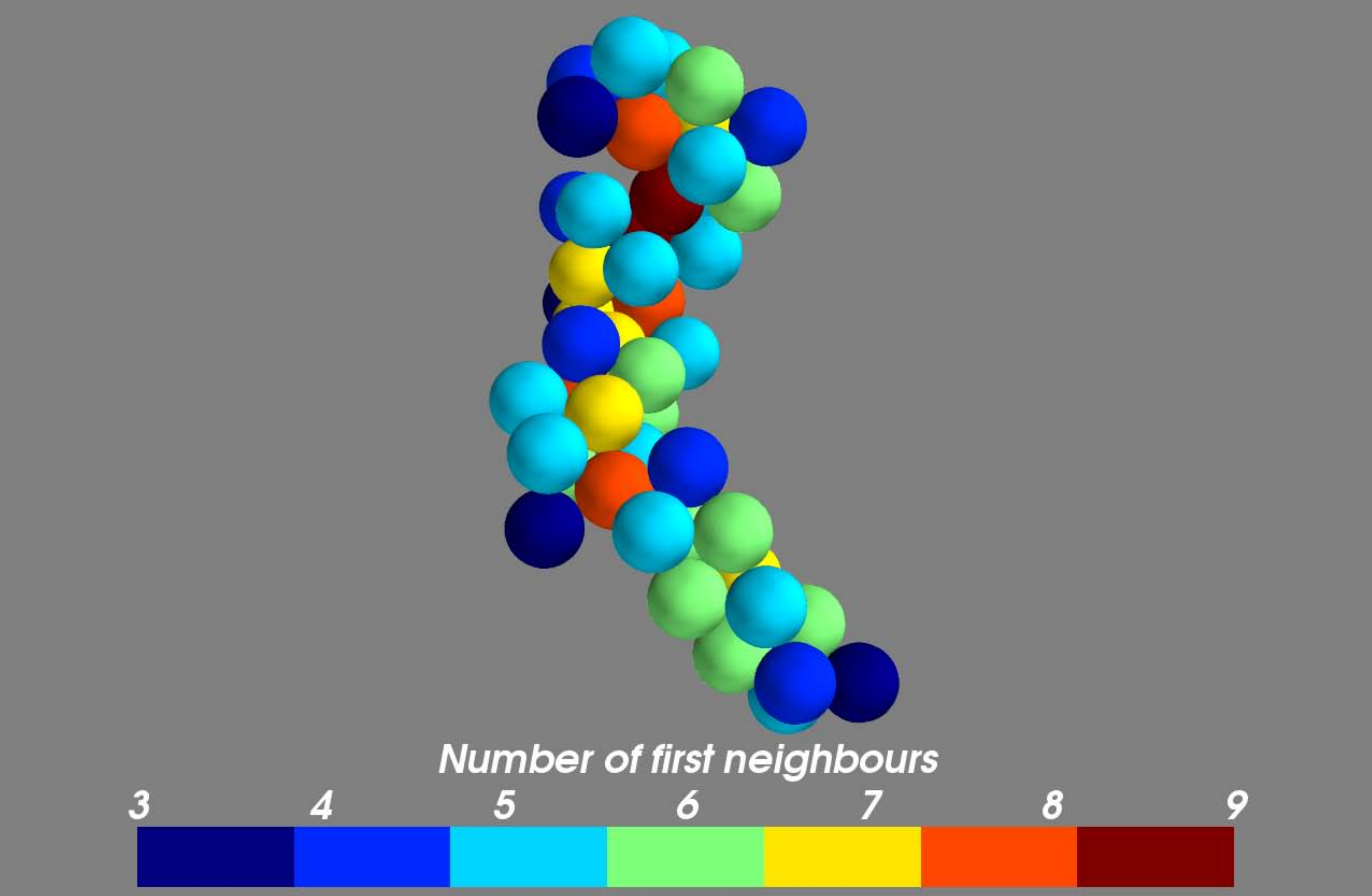}
\includegraphics[width=0.60\columnwidth,height=0.40\columnwidth]{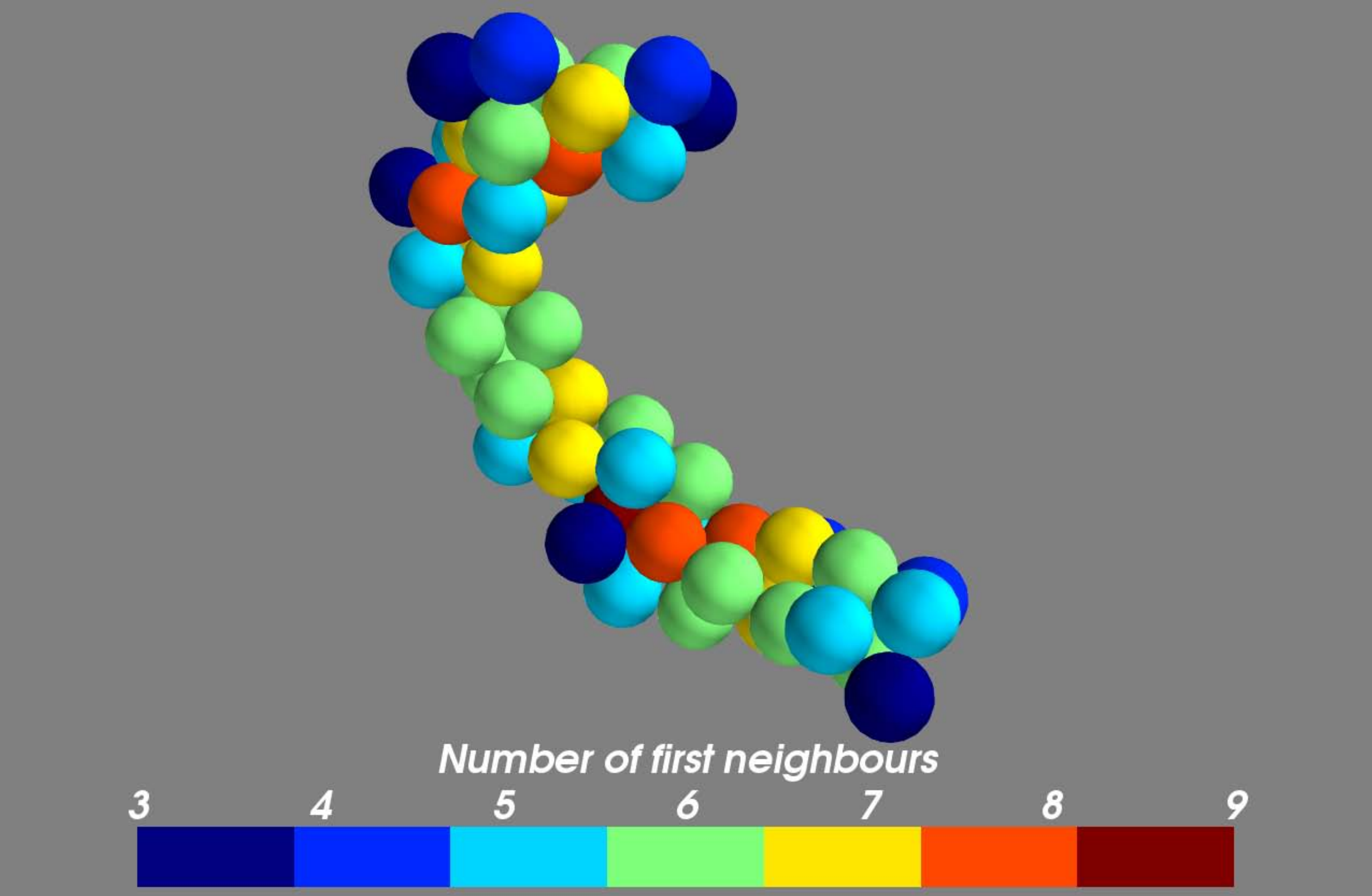}
\includegraphics[width=0.60\columnwidth,height=0.40\columnwidth]{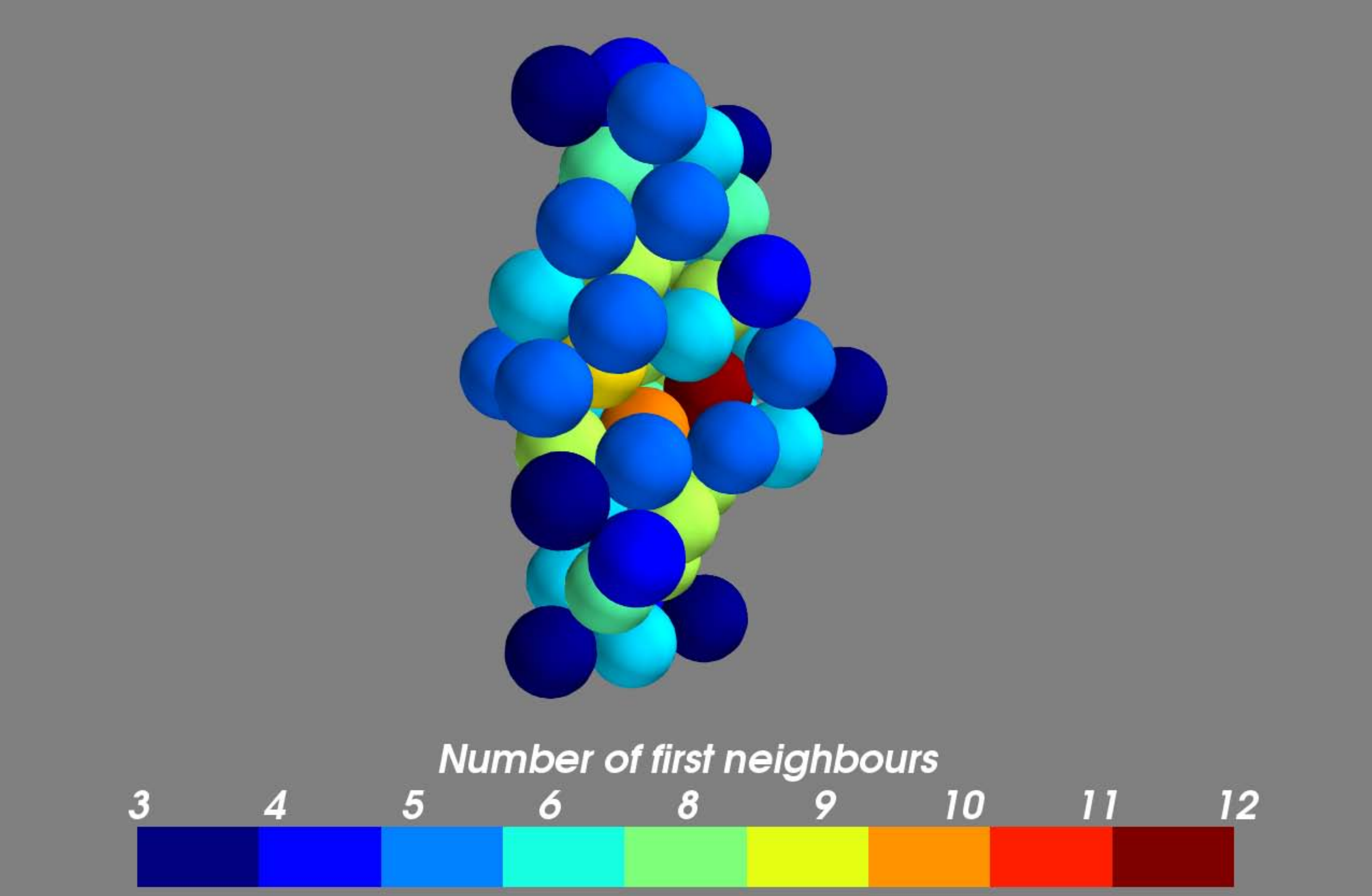}
\caption{Clockwise from top: Snapshot of the system at the end of a
simulation ($t_{\textrm{final}} = 3000$) followed by three different
$50$-monomer aggregates. Three dimensional images created with the
software Mayavi2~\cite{mayavi2}. The color code denotes number of first monomer neighbours.}
\label{snapshot_end_simulation}
\end{figure*}
\end{turnpage}

The local, small-scale compactness of the aggregates, aggregates that
are similar to those generated by the Langevin simulations of, e.g,
Ref.~\cite{videcoq}, is related to the spherical symmetry of the monomer-monomer interaction potential.
At early times when a monomer collides and binds to an aggregate
it is free to slide, a motion induced by the thermal noise, and to reorient to maximize the number
of contacts with other monomers. Since the potential is isotropic there is no
angular restoring force to prevent sliding. However, short-time restructuring is
hindered as the number of monomer-monomer contact points increase because the energy
cost associated with stretching monomer-monomer bonds is high.
Inspection of Fig.~\ref{snapshot_end_simulation} shows that the minimum
number of first neighrours in a stable cluster is three, suggesting that in three
dimensions three contact points are sufficient to prevent monomer sliding. Hence,
the aggregates generated herein are different from
those generated by diffusion-limited aggregation where colliding monomers remain fixed at the
initial point of contact. Becker and Briesen~\cite{TangentialForces}
showed that in the absence of a potential to prevent bending
of monomer-monomer bonds the aggregate collapses to a more compact
structure.
The monomer reorientation at early times also suggests that the
high fractal dimension $d_f^{mc}$ we associated with
monomer-cluster agglomeration mechanism arises from monomer
rearrangement. In our simulations monomer reorientation occurred
at very small time scales, smaller than the sampling time.
After the short-time monomer reorientation clusters remain
rigid till the next collision, as discussed Sec.~\ref{sec:clust-restructuring}.

An additional explanation for the compactness of
aggregates generated by Langevin simulations,
neglecting Brownian motion, was suggested by Tanaka
and Araki~\cite{prl_hydro}. They argued
that in the absence of interparticle hydrodynamic interactions,
in particular of squeezed-flow effects, the generated clusters
tend to be more compact.

Even though the clusters are locally compact, the large clusters generated at
late times have a low fractal dimension $d_f^{cc}<2$.
Inspection of Fig.~\ref{snapshot_end_simulation} (second and third clusters)
suggests that the low, late-time fractal dimension is due to their
tubular, elongated shape. The aggregates are not porous: they do not have holes nor
cavities. The large-scale structure of the late-time (large) clusters is
determined by the attractive range of the intermonomer potential.
At late times when two (locally compact) clusters collide aggregate
restructuring is limited to the monomers that are in contact:
the attractive range of the potentials used in the simulation is too short to
induce significant cluster re-organization, a process that would
increase the fractal dimension. Videcoq et al.~\cite{videcoq} showed that
increasing the attractive range modifies the shape of the generated aggregates
leading to more spherical aggregates. Hence, the late-time
fractal dimension associated with cluster-cluster aggregation depends
on the extend of restructuring due to the attractive potential range. This
restructuring provides a possible explanation of the lower
than expected fractal dimension for cluster-cluster aggregation.

Additional simulations were performed to assess the sensitivity of the observed
cluster morphology, and their rigidity, on the simulation time step.
We simulated a $50$-monomer cluster (shown at the bottom right, Fig.~\ref{snapshot_end_simulation})
for $32 000$ time steps ($t_{\textrm{final}} =40$), with two different
time steps, $\delta t_{\textrm{sim}}$ and $\delta t_{\textrm{sim}}/4$. We monitored the
time evolution of the cluster radius of gyration $R_g(t)$, an average cluster property, and
an instantaneous cluster property, the
distance $d_{1,12} (t)$ of two randomly chosen monomers
(monomers $1$ and $12$). We found that both quantities fluctuated by
approximately $0.1 \%$, independently of time step. These results confirm that
cluster morphology does not depend on the choice of the time step, as long as it is
chosen within reason. In fact, for considerably longer time steps the cluster breaks up.
The fluctuations arise because monomer radial positions fluctuate, albeit slightly, about the
potential minima.

As a different measure of a possible dependence of cluster
morphology on the simulation time step we compared the
time-dependent adjacency matrices, Eq.~(\ref{eq:adjacency_matrix}),
for these two simulations. The adjacency matrices were found to be
all identical, another confirmation that cluster morphology is
independent of the time step. Note that small fluctuations of
monomer radial positions in the cluster are not reflected in the
adjacency matrix (a binary matrix) since a distance threshold is
used to convert the distance matrix, Eq.~(\ref{eq:distance_3D}), to
it.

Thus, the morphology of the generated aggregates arises
from a combination of kinetic effects, as induced by the thermal
noise, and energetic effects, as determined by the attractive
range of the interaction potential. A precise characterization
of aggregate morphology requires both the fractal dimension and
the coordination number (or the lacunarity).


\subsection{Cluster restructuring}
\label{sec:clust-restructuring}

In addition to short-time monomer reorientation cluster restructuring
may occur following a cluster-cluster collision.
Cluster restructuring following contact of two clusters
was investigated in a three-dimensional
lattice model in Ref.~\cite{meakin_reorganization_3D}.
In our approach cluster restructuring does not have to be imposed as an
additional feature of cluster dynamics, but it occurs naturally.
The deep \textit{radial} intermonomer potential locks distances
between neighbouring monomers, but bonded monomers may slide over
each other due to the thermal noise term. However, as mentioned
earlier, monomer sliding is restricted by the number of
monomer-monomer contact points.

\begin{figure*}
\includegraphics[width=0.60\columnwidth]{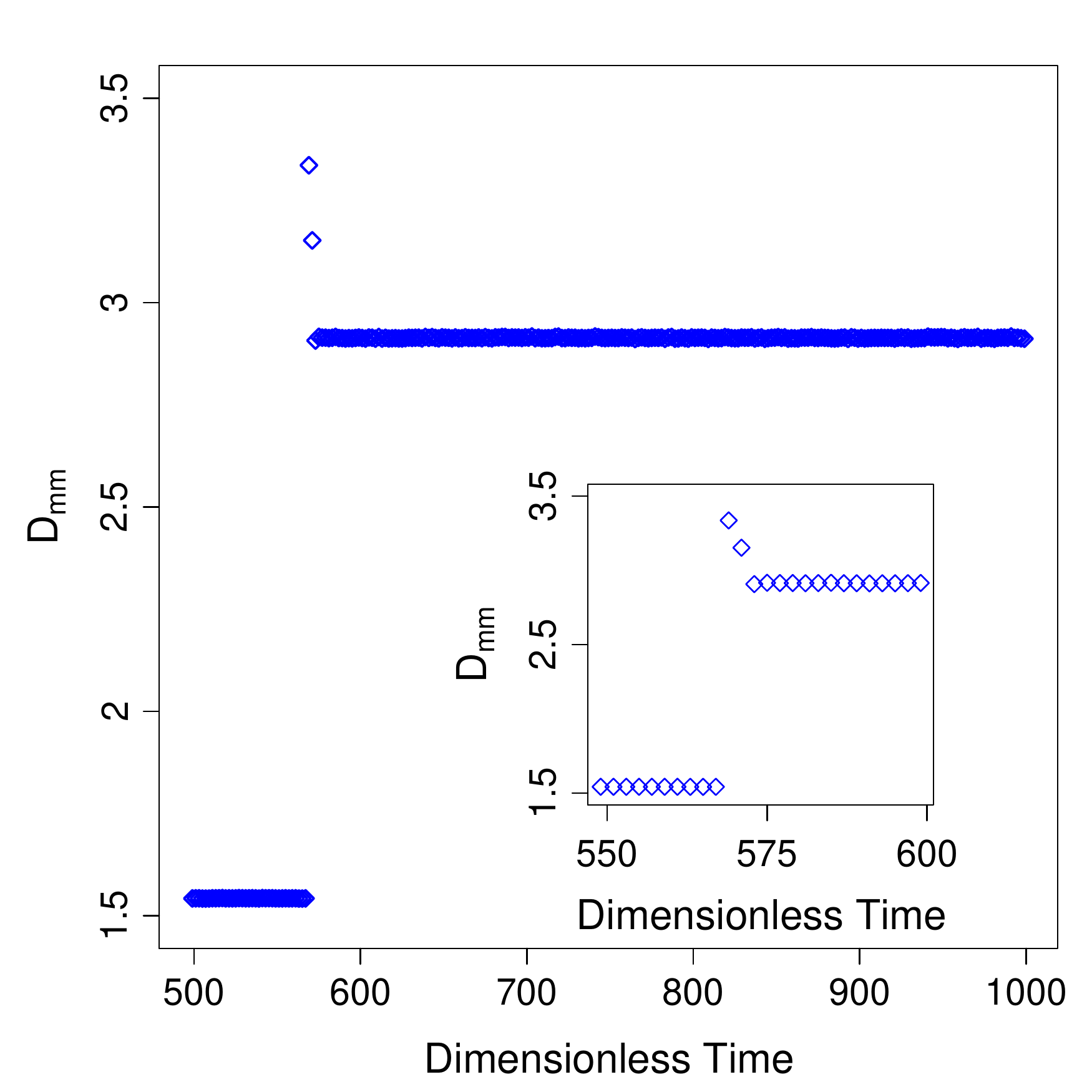}
\includegraphics[width=0.60\columnwidth]{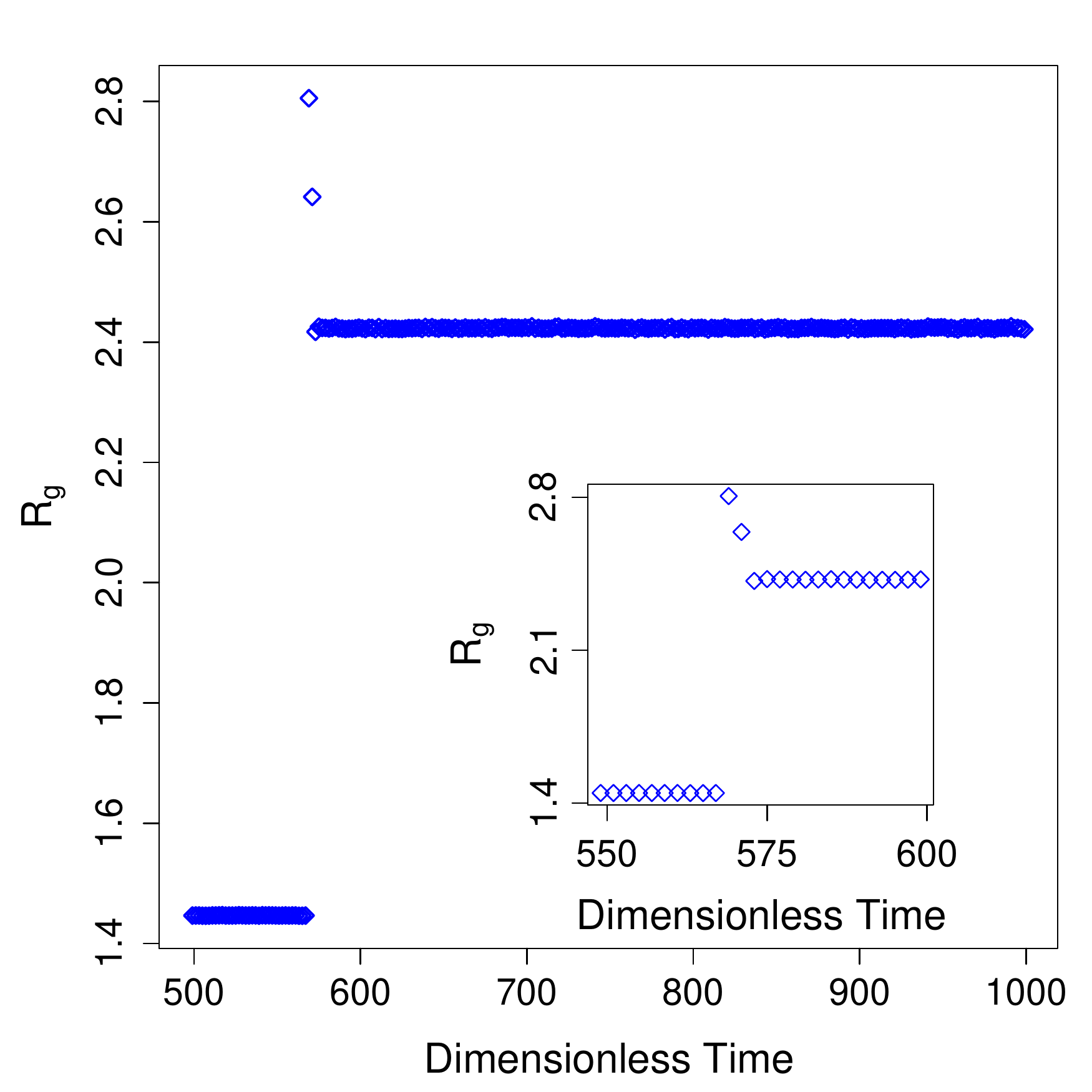}
\caption{Top: Mean monomer-monomer distance of a selected cluster.
Bottom: Radius of gyration of the same cluster.}
\label{single_cluster_size_evolution}
\end{figure*}

We use two statistical quantities as indicators of the modification of
the structure of a cluster: the radius of gyration and the mean
monomer-monomer distance. For a $k$-mer the mean monomer-monomer distance
$D_{mm}$ is the average monomer-monomer distance over all the monomers,
\beq
D_{mm} = \frac{1}{k^2} \, \sum_{i,j}^k D_{ij} \Comma
\eeq
where the monomer-monomer distance $D_{ij}$ was defined in Eq.~(\ref{eq:distance_3D}).
We monitored cluster restructuring by selecting a monomer at the beginning of the simulation
and follow it as it collides with other monomers or clusters.
In Fig.~\ref{single_cluster_size_evolution} (left) we show $D_{mm}$
during a period when a collision occurs ($t=568$) and the cluster size
increases from $k=14$ to $k=43$ . We notice that before and after
the collision $D_{mm}$ is constant,
a clear sign of cluster rigidity. When the collision between the two aggregates occurs,
$D_{mm}$ increases, but on a time scale of a few monomer relaxation
times it decreases, afterwards remaining relatively constant.
We consider this behaviour a strong indication of cluster reorganization
following the collision of two clusters. The cluster radius of gyration
shows a perfectly analogue behavior, Fig.~\ref{single_cluster_size_evolution} (right).
Note, however, that once cluster restructuring occurs after
a cluster-cluster collision the resulting cluster retains its
shape and it remains rigid until possibly the next collision.

\section{Dynamic cluster properties}
\label{sec:Dynamics}


\subsection{Cluster translational diffusion coefficient}
\label{sec:clust-diff-coeff}

Equations~(\ref{eq:Langevin}) determine the dynamics of each monomer and indirectly
the cluster diffusive properties.
The cluster diffusion coefficient is calculated
from the late-time dependence of the
variance of the cluster center-of-mass position as a function of
time~\cite{risken_book}
\begin{equation}
\label{eq:diffusion_simu_cluster}
\langle \delta{ R}^2_{CM}(t) \rangle =
\langle \big [ {\bf R}_{CM} (t) - \langle {\bf R}_{CM} (t) \rangle \big ]^2 \rangle \
\xrightarrow{t\to\infty} 6D_{k}t \Teleia
\end{equation}
A different set of simulations was performed to determine
the diffusion coefficient of a few selected clusters. As noted in
Sec.~\ref{sec:clust-restructuring} in the absence of collisions
clusters are rigid: the dynamical properties presented in this
section refer to perfectly rigid clusters. A selected cluster
was placed in the simulation box with its centre-of-mass
position at ${\bf R}_{CM}=(0,0,0)$ and with zero monomer velocities.
The box size was chosen to be $L=10,000$,
a size much larger than the box size used for the agglomeration simulations.
A large box is necessary
because the aggregate centre of mass has to be tracked for a (potentially) long time,
and, due to the periodic boundary conditions, no displacement of the aggregate from its
initial position larger than $L/2$ is admissible.
For these simulations the cluster never crossed the boundary of the box.

\begin{figure*}
\includegraphics[width=0.60\columnwidth]{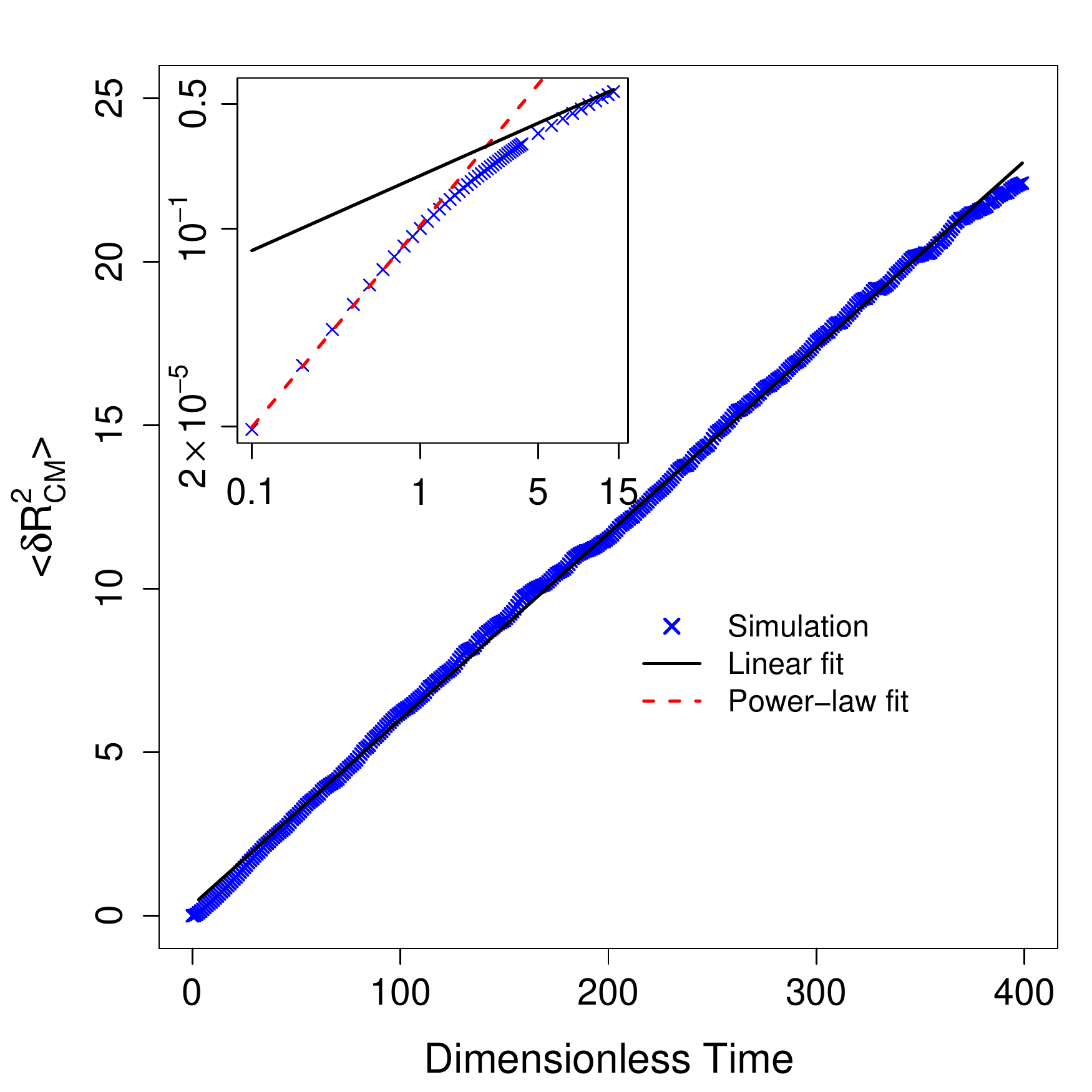}
\includegraphics[width=0.60\columnwidth]{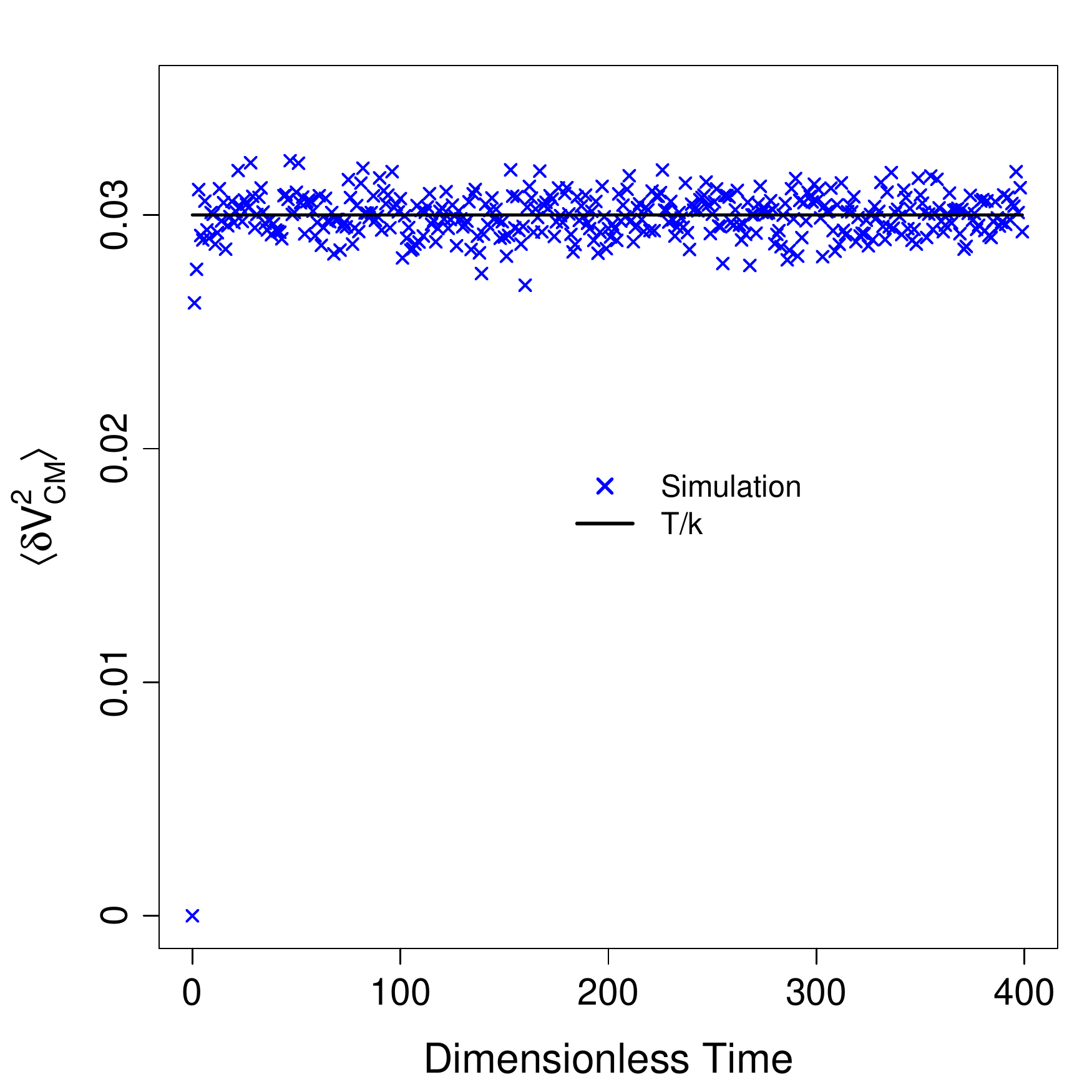}
\caption{Top: Mean-square cluster displacement averaged over
800 trajectories (crosses) and linear fit (solid line) of a $50$-monomer
cluster (shown in Fig.~\ref{snapshot_end_simulation}, bottom right).
Inset: Early-time behaviour of $\langle \delta R^2_{CM}  \rangle$ (crosses),
linear fit (solid line) and power-law fit, $\gamma =3$ (dashed line).
Bottom: Velocity fluctuations $\langle \delta V^2_{CM}  \rangle$
(crosses) and analytical expression, Eq.~(\ref{eq:delta_v_squared}) (solid line).}
\label{diffusion_plot}
\end{figure*}

The center of mass of a single cluster was tracked up to $t=400$ for clusters
with $k=4,10,50,98$. Averages were performed over $800$ trajectories, each trajectory
starting with an identical cluster: different cluster trajectories
arise from different realizations of the
stochastic noise. Figure~\ref{diffusion_plot} (top) shows $\langle \delta{R}^2_{CM}(t) \rangle$
for a $50$-monomer cluster,
chosen to be the cluster at the bottom right of Fig.~\ref{snapshot_end_simulation}.
The time dependence of the ensemble-averaged, mean-square cluster displacement quickly becomes linear,
in agreement with Eq.~(\ref{eq:diffusion_simu_cluster}).
The inset of Fig.~\ref{diffusion_plot} (top) magnifies (on a double
logarithmic scale) the early-time behavior of
$\langle \delta{R}^2_{CM}(t) \rangle$. For $t\le 1$ cluster motion
is ballistic, and the mean-square displacement
exhibits a power-law dependence on time,
$\langle \delta{R}^2_{CM}(t) \rangle\propto t^\gamma$ with $\gamma = 3$.
This is the expected behaviour for a single Brownian
monomer with zero initial velocity \cite{risken_book}.
Similar early-time ballistic motion was found for a cluster
with $k=4$.

The results of the numerical simulations for the four clusters
are summarized in Table~\ref{table_diffusion},
where we also report the cluster radius of gyration
(obtained from Fig.~\ref{two_df}).
They may be interpreted by considering
theoretical estimates of the diffusion coefficient $D_k$ of a $k$-mer.
The Stokes-Einstein diffusion coefficient of a
single spherical monomer, in dimensionless
form, scaled by $D^{\star} = k_B T^{\star} /( \beta _1 m_1)$, is \cite{risken_book}
\begin{equation}
\label{eq:monomer-diffusion-coefficient}
D_1=  T \Teleia
\end{equation}
The diffusion coefficient of a $k$-mer of mass $M_k = k m_1$ may be
expressed as a generalization of Eq.~(\ref{eq:monomer-diffusion-coefficient}).
Let $\beta_k$ be the average friction coefficient of a
monomer in the $k$-cluster. Then, the cluster drag term in Eq.~(\ref{eq:Langevin})
may be written as $\beta_k M_k = k \beta_k m_1$, and the cluster diffusion coefficient generalizes
to
\begin{equation}
\label{eq:cluster-diffusion-coefficient}
D_k = D_1 \, \frac{\beta_1}{k \beta_k}
\equiv D_1 \, \frac{1}{k \eta_k} \Teleia
\end{equation}
Equation~(\ref{eq:cluster-diffusion-coefficient}) manifestly shows
the importance of the ratio of the average friction coefficients
$\eta_k = \beta_k / \beta_1$, $\eta_k$ being the average shielding factor
of a monomer in a $k$-cluster. In fact,
\citet{filippov_thermal}, in a different context, used the cluster
shielding factor to describe how monomer shielding affects
heat transfer to an aggregate.
Herein, the cluster shielding factor is connected to cluster
mobility.

The numerically determined diffusion coefficients behave as $D_k\propto 1/k$
with $k$ spanning almost two orders of magnitude. 
The estimated value of $\beta_k$ equals the monomer friction coefficient
$\beta_1$ within a few percents,
Table~\ref{table_diffusion} ($\beta_1 =1$ in our units). The small deviations are
attributable to the limited number of
simulated stochastic trajectories. Our Langevin simulations suggest
that the cluster diffusion coefficient is inversely proportional to the cluster
mass and to the monomer friction coefficient. This behaviour is a direct
consequence of neglecting shielding of inner monomer by outer monomers in an
aggregate. The cluster accessible area becomes the sum of the
monomer accessible areas, since for $\eta_k =1$ the total friction
coefficient of a $k$-mer is $3 \pi \mu_f \sigma k$.
For this reason we refer to these
clusters as \textit{ideal} clusters, with respect to their dynamical properties.
For ideal clusters the average monomer shielding factor is unity irrespective
of the state of aggregation of the monomer. This is an inherent property
of all simulations of monomer agglomeration via the unshielded
monomer Langevin equations Eqs.~(\ref{eq:Langevin}).

The cluster diffusion coefficient (in the continuum regime) is frequently expressed
in terms of a mobility radius $R_m$, the radius of a sphere with the same
friction coefficient as the aggregate \cite{friedlander book}. In the dimensionless units
used in this work,
the mobility radius determines the cluster diffusion coefficient by
\begin{equation}
\label{eq:mobility_radius}
D_k = D_1 \, \frac{1}{2 R_m} \Teleia
\end{equation}
Comparison of Eqs.~(\ref{eq:cluster-diffusion-coefficient}) and (\ref{eq:mobility_radius})
gives an expression for the mobility radius in terms of the (average) cluster (per-unit-mass) friction
coefficients $\beta_k$ and $\beta_1$, or the shielding factor $\eta_k$,
\beq
R_m = \frac{k}{2} \, \frac{\beta_k}{\beta_1} = \frac{k}{2} \, \eta_k \Teleia
\label{eq:DiffusionShielding}
\eeq
For the ideal clusters generated in this work $R_m = k/2$.
Table~\ref{table_diffusion} presents the mobility radius calculated either
from the numerically determined diffusion coefficient [Eq.~(\ref{eq:mobility_radius}),
fourth row], or by evaluating it directly from Eq.~(\ref{eq:DiffusionShielding}) with $\eta_k =1$ (fifth row).
A comparison of these two values provides
an indication of the numerical error of our simulations: the percentage
shown in the fifth row (in parenthesis) quantifies the difference.

The cluster diffusion coefficient is frequently estimated by
taking the aggregate mobility radius $R_m$ approximately
equal to the cluster radius of gyration $R_g$.
Our simulations
show that the radius of
gyration is of the same order of magnitude for cluster sizes in the range
$k=4$ to $k=98$, whereas the mobility radius is considerably higher.
Since the radius of gyration severely underestimates the mobility radius,
the diffusion coefficient of ideal clusters is lower
than the diffusion coefficient of clusters with $R_m \sim R_g$.
In ideal clusters the average monomer shielding factor is unity,
whereas monomers in non-ideal clusters are at least partially shielded, resulting in higher
cluster mobility, cf. Eq.~(\ref{eq:cluster-diffusion-coefficient}).

\subsection{Cluster thermalization}
\label{sec:Thermalization}

For a single Brownian monomer, the
late-time mean-square velocity fluctuations (dimensionless, scaled
by $k_B T^{\star}/m_1$) tend to \cite{risken_book}
\begin{equation}
\label{eq:delta_v_squared_monomer}
\langle \delta v_{1, \infty}^2 \rangle =
\lim_{t \rightarrow \infty} \langle \big [ {\bf v}_1(t) - \langle {\bf v}_1 (t) \rangle \big ]^2 \rangle
 = T  \Comma
\end{equation}
where ${\bf v}_1(t)$ is the monomer instantaneous velocity. For a $k$-mer the
previous expression, a manifestation of energy equipartition
and a consequence of the FDT, may be generalized to
\begin{equation}
\label{eq:delta_v_squared}
\langle \delta  V_{CM, \infty}^2 \rangle =
\lim_{t \rightarrow \infty} \langle \big [ {\bf V}_{CM} (t) - \langle
{\bf V}_{CM}(t) \rangle \big ]^2 \rangle
= \langle \delta v_{1, \infty}^2 \rangle \, \frac{1}{k} \Comma
\end{equation}
where the cluster centre-of-mass velocity is ${\bf V}_{CM} (t) = \sum_i {\bf v}_i/k$,
with ${\bf v}_i$ the $i$-th monomer velocity. As for a Brownian monomer
Eq.~(\ref{eq:delta_v_squared}) expresses energy equipartition for a $k$-monomer cluster.
In Fig.~\ref{diffusion_plot} (bottom) we plot $\langle \delta V_{CM}^2 (t) \rangle$
as a function of time for the 50-monomer cluster used in the previous section.
As shown, Fig.~\ref{diffusion_plot} (bottom), $\langle \delta V_{CM, \infty}^2 (t) \rangle$ tends
at late times to $0.03$ (up to noise fluctuations), i.e., the theoretical value for ideal clusters,
Eq.~(\ref{eq:delta_v_squared}), for the simulation parameters  $T=0.5$ and $k=50$.

The main result of Secs.~\ref{sec:clust-diff-coeff}
and~\ref{sec:Thermalization} is that a monomer in an ideal $k$-cluster
generated according to Eqs.~(\ref{eq:Langevin}) and (\ref{eq:gaussian_noise}),
i.e., neglecting monomer shielding, feels the same friction
coefficient as an isolated monomer ($\beta_k = \beta_1$).
An ideal $k$-monomer aggregate has the same diffusive properties as a
monomer sphere
of $k$ times the mass of a single monomer ($D_k = D_1/k$).
The cluster diffusion coefficient differs from the monomer
diffusion coefficient only due to the larger cluster mass, and not, in addition,
due to the decrease of the average monomer shielding factor.

\subsection{Cluster rotation}
\label{sec:clust-diff-rotation}

We investigated whether ideal clusters have a preferential orientation
by examining the distributions of their average Euler angles.
As argued, clusters in the absence of collisions may be treated as rigid bodies.
Rigid body rotation in three dimensions may be
described by the three Euler angles $\theta$, $\phi$ and $\psi$ \cite{goldstein}.
We evaluated them during post-processing
by eliminating cluster translational motion via rigidly
translating the cluster centre of mass to the origin of the
computational-box coordinate system at all times.
The Euler angles may, then, be calculated by recording
the time-dependent positions
${\bf r}_A (t)$, ${\bf r}_B (t)$ and ${\bf r}_C (t)$  of three non-coplanar monomers.
We define a 3 by 3 matrix ${\bf X}(0) =[{\bf r}_A (0), {\bf r}_B (0), {\bf r}_C (0)]$
containing the initial positions of the three reference monomers  and a
matrix ${\bf X}' (t) =[{\bf r}_A (t), {\bf r}_B (t), {\bf r}_C (t)]$ with
their positions at time $t$. The rotation matrix ${\bf A}$ such that
${\bf X}'= {\bf AX}$ is
\begin{equation}
\label{eq:rotation_matrix}
{\bf A} (t) = {\bf X}'(t) {\bf X}^{\textrm{T}} (0) \,
\Big [ {\bf X} (0) {\bf X}^{\textrm{T}} (0)  \Big ]^{-1} \Comma
\end{equation}
where the superscript T denotes matrix transpose.
Once the rotation matrix $\bf A$ is known,  the Euler angles may be determined.
Since they are not uniquely defined, their values depending on the order of the
three rotations, we chose to calculate them via the algorithm presented
in Ref.~\cite{rotation_alghoritm}.
The range of the Euler angles so determined is: $\phi$ and $\psi$
range in the interval $[-\pi,\pi]$, whereas $\theta$ lies in the interval $[-\pi/2,\pi/2]$.
For uniform random rotation matrices \cite{random_euler}, both $\psi$
and $\phi$ are random variables with a uniform probability distribution
in the interval $[-\pi,\pi]$:
hence $\langle\phi\rangle=\langle\psi\rangle=0$ and
$\langle \delta \phi^2(t) \rangle^{1/2} =
\langle [ \phi(t) - \langle \phi (t) \rangle ]^2 \rangle^{1/2} =
\langle \delta \psi^2 (t) \rangle^{1/2}  ={\pi}/{\sqrt3}\simeq 1.81$.
On the other hand, the Euler angle $\theta$ is distributed according to \cite{random_euler}
\begin{equation}
\label{eq:theta_distr}
\theta=\arccos(1-2Z(0,1))-\f{\pi}{2} \Comma
\end{equation}
where $Z(0,1)$ is a random variable with a uniform probability
distribution in the interval $[0,1]$. The corresponding averages,
obtained numerically, are
$\langle \theta (t)\rangle=0$ and
$\langle \delta \theta^2 (t) \rangle^{1/2} \simeq 0.68$.

\begin{figure*}
\includegraphics[width=0.50\columnwidth,height=0.50\columnwidth]{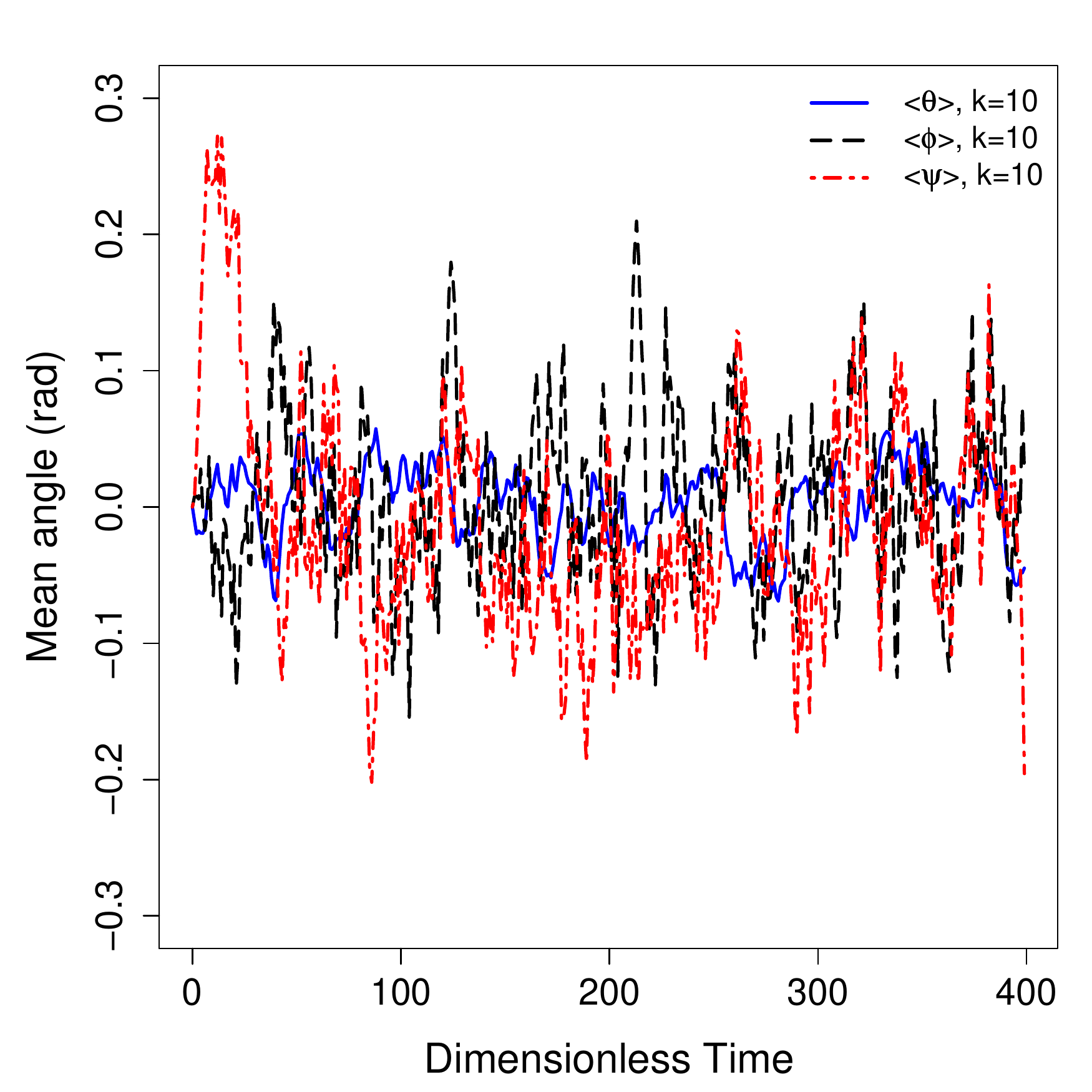}
\includegraphics[width=0.50\columnwidth,height=0.50\columnwidth]{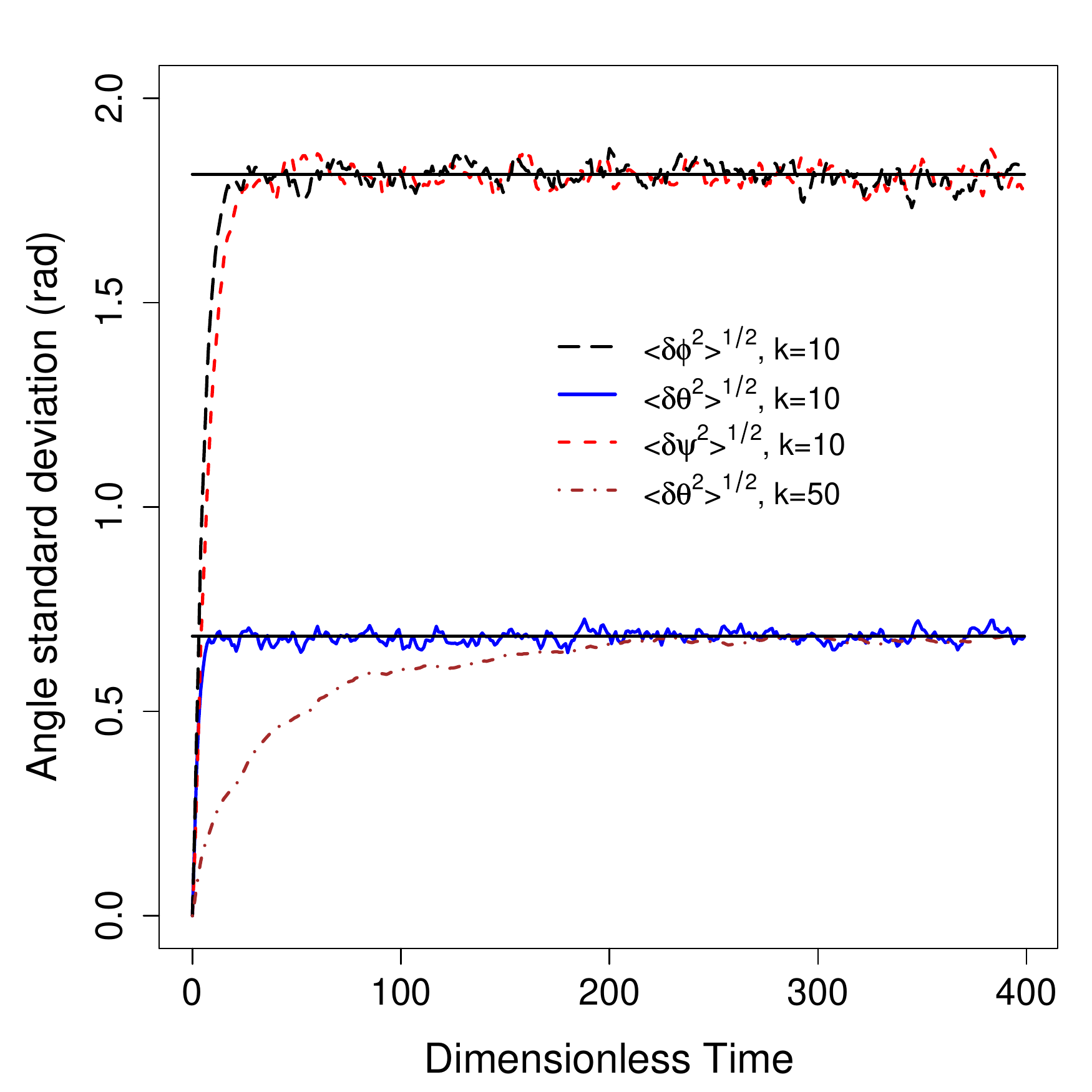}
\caption{Top: Mean Euler angles $\l\theta\r$ (solid line),
$\l\phi\r$ (dashed line) and $\l\psi\r$ (dot-dashed line), averaged over 800
trajectories, for a $10$-monomer cluster.
Bottom: Standard deviations
$\l\delta\theta^2\r^{1/2}$ (solid line), $\l\delta\phi^2\r^{1/2}$
(long-dashed line), and $\l\delta\psi^2\r^{1/2}$ (short-dashed line) for the
same $10$-monomer cluster, and $\l\delta\theta^2\r^{1/2}$
(dot-dashed line) for a $50$-monomer cluster.
The horizontal lines are the standard deviations of random rotation matrices.}
\label{cluster_rotation}
\end{figure*}

Figure~\ref{cluster_rotation} (top) shows the mean Euler angles
calculated for an ensemble of $800$ identical
clusters containing $10$ monomers each.
Their mean values fluctuate about zero.
On the bottom diagram we show their mean-square fluctuations:
they rapidly tend to the expected values
for random rotation matrices. Another calculation of
$\l\delta\theta^2\r$ for an ensemble of $800$ identical cluster with
$k=50$ shows saturation to the same value but on a longer time scale.
The rotational thermalization time of an ideal $k$-cluster
depends on its total mass.
Hence, at late times an ideal cluster does not have a
preferential orientation.

\subsection{Time dependence of cluster number: The agglomeration
equation}
\label{sec:AgglomerationEquation}

The instantaneous total number of clusters, $N_{\infty} (t) = \sum_k^{\infty} N_k(t)$,
quantifies the progress of agglomeration. It may be calculated
via the numerical solution of the agglomeration equation \cite{friedlander book}
\begin{equation}
\label{eq:agglomeration_equation}
\f{dn_k}{dt}=\f{1}{2}\s_{i+j=k} K_{ij}n_in_j-n_k\sum_i K_{ik}n_i
\Teleia
\end{equation}
For definitions and scalings see Section~\ref{sec:Kernel}. In this section
the collision kernel $K_{ij}$ is scaled by
$K^{\star} = D^{\star} \sigma = \beta_1 \sigma^3$, see Eq.~(\ref{eq:CharacteristicTemperature}).

The standard treatment of the collision kernel for fractal aggregates
in the continuum regime gives \cite{friedlander book}
\begin{equation}
\label{eq:cont_kernel_general}
K_{ij}=4 \pi (D_i+D_j)(R_{i}+R_{j}) \Comma
\end{equation}
where we dropped the subscript $g$ in the radius of
gyration of an $i$-mer, $R_i=R_{g,i}$, to simplify notation.
Equation~(\ref{eq:cont_kernel_general}) is
often expressed in terms of the aggregate volume of solids
$\nu_i\propto(R_{i})^{d_f}$ \cite{LorenzoJAS}
since the volume of solids is conserved
during agglomeration. Moreover, the diffusion coefficients in Eq.~(\ref{eq:cont_kernel_general})
are given by Eq.~(\ref{eq:mobility_radius}) with the
mobility radius being the radius of gyration. The kernel then reads
\begin{equation}
\label{eq:cont_kernel_standard}
K_{ij}^{Sm} = 2 \, \pi \, D_1 \,
\lro \f{1}{\nu_i^{1/d_f}}+\f{1}{\nu_j^{1/d_f}}  \rro \,
\lro \nu_i^{1/d_f}+\nu_j^{1/d_f} \rro \, ,
\end{equation}
where the superscript $Sm$ refers to the so-called Smoluchowski kernel.
The Smoluchowski kernel is not expected to model our numerical
results accurately because it is based on diffusion coefficients given by
Eq.~(\ref{eq:mobility_radius}) with $R_m=R_g$, an equality not
respected by our simulations.
Instead, the numerical simulations may be used to derive a
modified kernel appropriate for the reproduction of the numerical results.
Such a kernel is obtained by
using the numerically determined cluster
diffusion coefficients, Eq.~(\ref{eq:cluster-diffusion-coefficient})
with $\beta_k=\beta_1$ in Eq.~(\ref{eq:cont_kernel_general})
For the radius of gyration, we use the fitted value for
$k\ge 5$ (and the data reported in Table~\ref{tab:RadiusScaling})
to obtain
\begin{equation}
\label{eq:rules_for_r_gyr}
R_i = \begin{cases}
\textrm{simulation data}, & i< 5 \Comma \\
a_{mc} \, i^{1/d_f^{mc}},  & 5\le i\le 15 \Comma \\
a_{cc} \, i^{1/d_f^{cc}}, &  15 < i \Teleia \\
\end{cases}
\end{equation}
Furthermore, non-continuum effects arising from
monomer-monomer collisions, and dependent on the
monomer mean free path, have been shown to
be important in combustion-generated nanoparticle
agglomeration~\cite{LorenzoJAS}. They
may be accounted for by introducing the (dimensionless) Fuchs
correction factor $\beta_F$ in the kernel: for the explicit
expression of $\beta_F$ consult Ref.~\cite{LorenzoJAS}.
Hence, the kernel appropriate for the Langevin simulations (hereafter
referred to as Langevin-Dynamics kernel) reads
\begin{equation}
\label{eq:kernel_complete}
K_{ij}^{LD} = 4 \, \pi \, D_1 \,
\lro\f{1}{i}+\f{1}{j}\rro \lro \,
R_{i}+R_{j}\rro \, \beta_F \Teleia
\end{equation}

The late-time dependence of the total cluster number may be
expressed in terms of the kernel homogeneity exponent $\lambda$
\cite{ErnstvanDongen87}, whereby a
kernel $K_{ij}$ is a homogeneous function of
order $\lambda$ if $K_{\gamma i,\gamma j}=\gamma^{\lambda} K_{ij}$.
Then, the asymptotic time-decay of the total cluster number
is $N_\infty (t) \sim t^{-1/(1-\lambda)}$.
For the Smoluchowski kernel $\lambda=0$ and, therefore, $N_\infty\sim t^{-1}$,
whereas for the Langevin-Dynamics kernel $\lambda=(1/d_f^{cc})-1$, leading to
$N_{\infty}\sim t^{-0.74}$.

The results of the simulations (with both potentials) for the
total number of clusters and the numerical
solution of the agglomeration equation
with $K^{Sm}$ and $K^{LD}$ are shown in Fig.~\ref{Smoluchowsky  comparison}.
The numerical solution of the agglomeration equation
with the Langevin-Dynamics kernel (short dashes)
shows good agreement with the simulations (dots, simulated intermonomer potential)
The early-time agreement is attributed to the Fuchs correction factor,
an observation also made in Ref.~\cite{LorenzoJAS}.

The time decay of the total cluster number was fitted
to a power law $N_\infty\sim t^{-\xi}$ at
times $ 2500 \le t\le 3000$. For the Langevin-Dynamics kernel we find
$\xi=0.77$, a value close to the exponent determined
from the numerical simulations ($\xi=0.78$ and $\xi=0.79$ for
model and the simulated potentials, respectively) and to $\xi =0.74$ expected
for homogeneous kernels. The
Smoluchowski kernel exponent is considerably different, $\xi=1$.
The slight difference between calculated and theoretical exponents
is attributed to the slow approach to the asymptotic limit of
the agglomeration equation: the asymptotic limit is reached
at times about one order of magnitude longer than the
duration of the simulations, primarily due to the Fuchs factor.
Finally, we note that, as expected, the Van der Waals attraction
enhances the agglomeration rate (compare the simulation results
with the two potentials), as noted in Ref.~\cite{videcoq}.

\subsection{Collision kernel elements}
\label{sec:KernelElementCalculate}

\begin{figure*}
\includegraphics[width=\columnwidth]{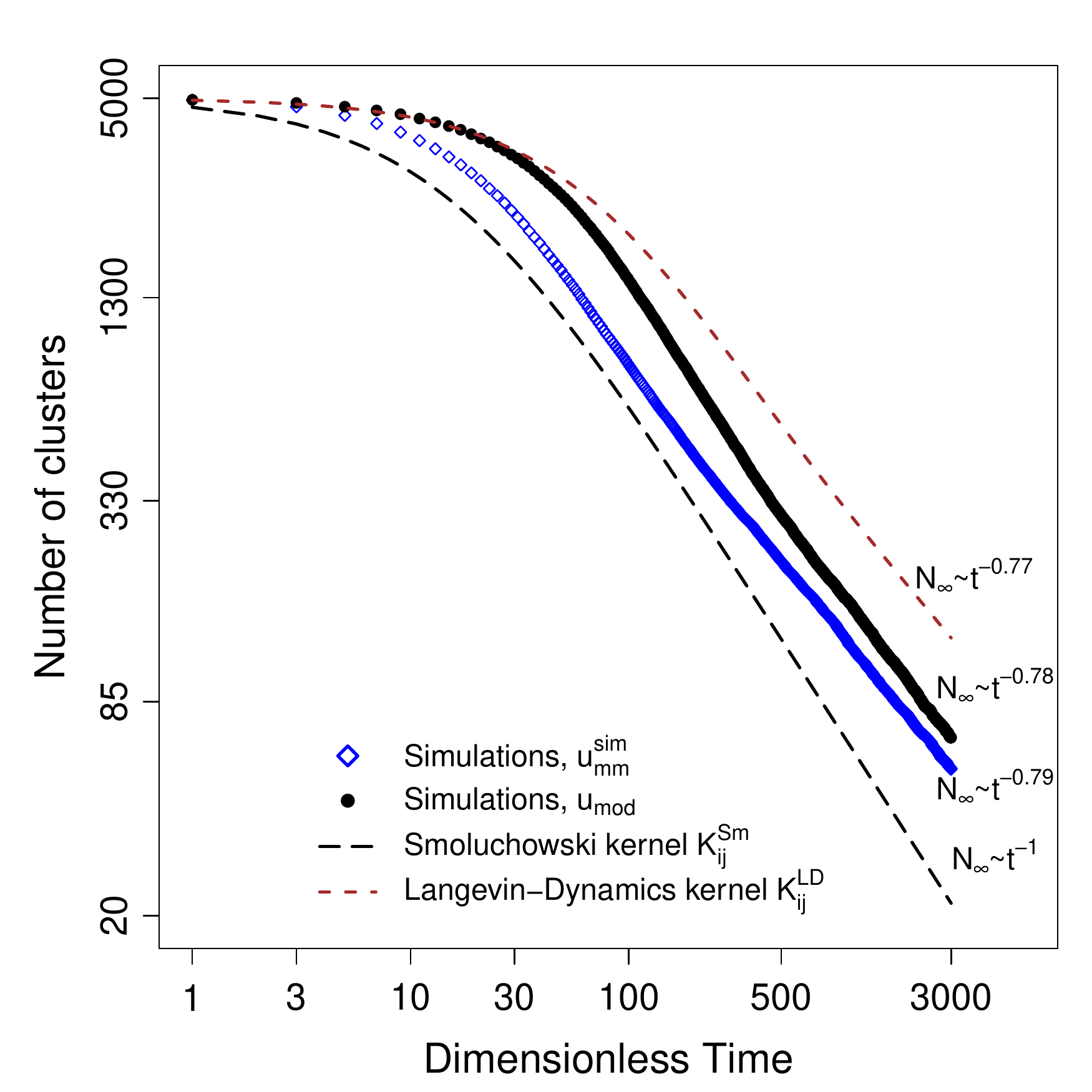}
\caption{Total number of clusters as a function of time calculated with
the simulated intermonomer potential (diamonds), model potential (filled circles),
numerical solution of the agglomeration equation with the Smoluchowski kernel
(long-dashed line) and with Langevin-Dynamics kernel (short-dashed line).}
\label{Smoluchowsky comparison}
\end{figure*}

\begin{figure*}
\includegraphics[width=\columnwidth]{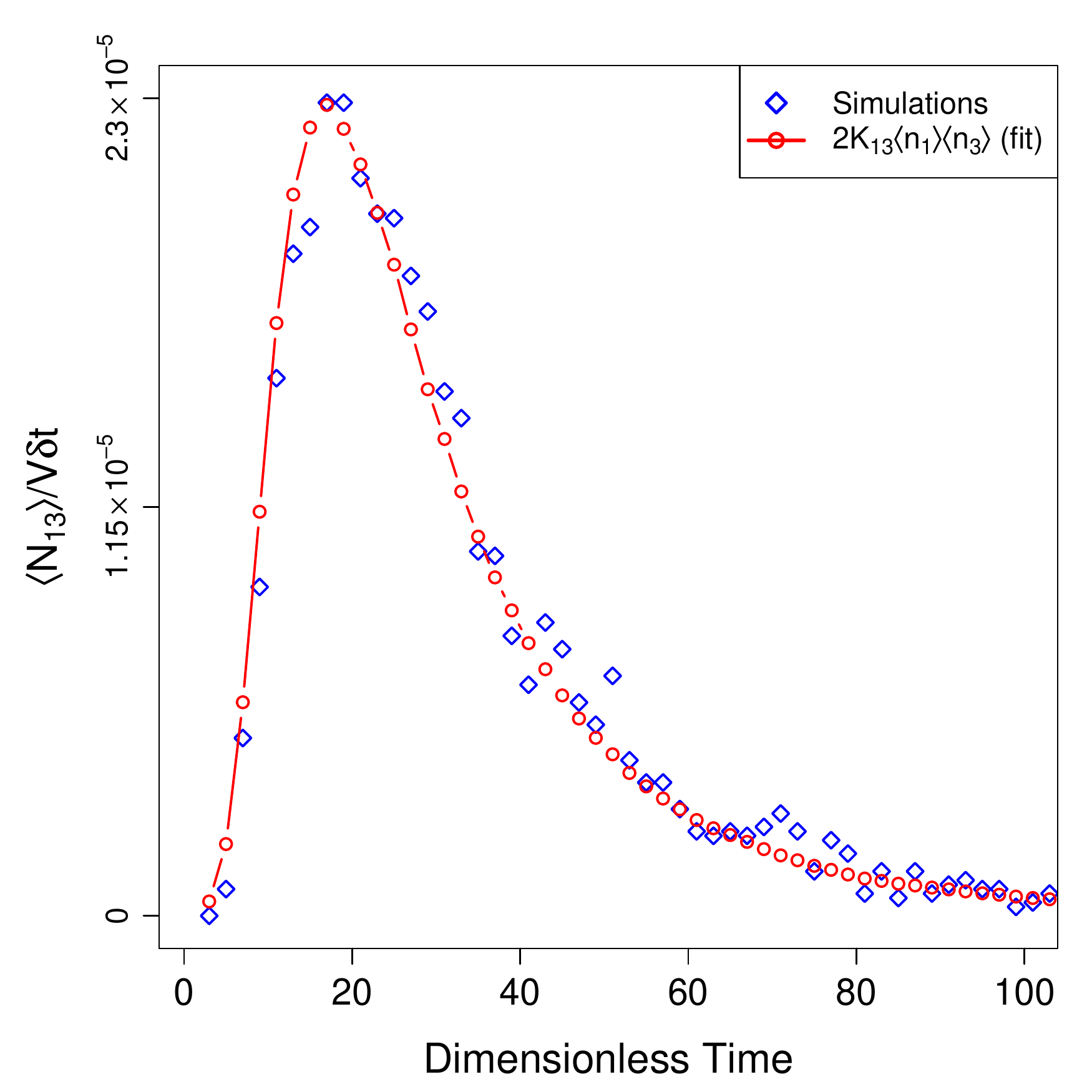}
\caption{Determination of kernel element  $K_{13}$: $\langle N_{13} \rangle /\delta t$ diamonds,
$2K_{13} \langle n_1\rangle \langle n_3 \rangle$ with the fitted
value of $K_{13}$ circles.}
\label{calculated_beta_ij}
\end{figure*}

Collision kernel elements may be obtained
by ensemble averaging Eq.~(\ref{eq:kernel_calc}) over the 10
initial conditions, i.e.,
\begin{equation}
\label{eq:kernel_nonlin_aver}
\f{\langle N_{ij}(t)\rangle}{V \delta t_{\textrm{sam}}}=
(2-\delta_{ij}) K_{ij}\langle n_i(t)n_j(t) \rangle \Teleia
\end{equation}
We found that the average of the nonlinear term decouples,
$\langle n_in_j\rangle = \langle n_i\rangle \langle n_j\rangle$,
to the accuracy of our simulations. We recorded cluster collisions
every $\delta t_{\textrm{sam}}$, to avoid counting ternary
collisions. We had enough data to
calculate only a few kernel elements
$K_{ij}$ for low $i$ and $j$ indexes.
Figure~\ref{calculated_beta_ij} shows an example of the fitting
procedure to determine $K_{13}$ from $\langle N_{13} \rangle /(V \delta t_{\textrm{sam}})$ and
$\langle n_1 \rangle \langle n_3 \rangle$.
The calculated kernel elements $K_{ij}$ are reported in
Table~\ref{table_kernel}, and they are compared to the
analytical values of $K_{ij}^{LD}$. Kernel elements are scaled
to the diagonal of the Smoluchowski kernel $K^{Sm}_{ii} = 8 \pi D_1$.
We note an excellent agreement for the $\{(1,1), (1,2), (2,2)\}$
elements. These kernel elements are important in the early-time
behaviour of $N_\infty(t)$, and therefore they are responsible for
the agreement of the numerical simulations with the numerical
solution of the agglomeration equation at early times, as shown
in Fig.~\ref{Smoluchowsky comparison}.

\section{Summary and conclusions}
\label{conclusions}

The static structure and diffusive properties of
aggregates formed in a quiescent fluid by the collision, and subsequent binding, of spherical
monomers were investigated via Langevin dynamics, in the limit
of small Knudsen number (continuum regime). The Langevin
equations of motion of a collection of diffusing and interacting
monomers were solved numerically using a package for
generic molecular dynamics simulations with a Langevin thermostat.
Two intermonomer interaction potentials were used: a model potential
and a potential that arises from the integration of the Lennard-Jones
intermolecular potential over the volume of two monomers. Both potentials
are spherically symmetric and rapidly decaying. Aggregates
were identified during post-processing by considering them to be
the connected components of a non-directed graph.

The static structure of the generated aggregates was described
in terms of their average fractal dimension and cluster coordination number.
We found that the aggregate fractal dimension varied with time
from an early-time value characteristic of monomer-cluster agglomeration,
($d_f = 2.25 \pm 0.05$), as determined by local monomer rearrangement and
restructuring,
to a late-time value characteristic of cluster-cluster agglomeration
($d_f = 1.56 \pm 0.02$), a value dependent on the attractive range of the
intermonomer potential. The time dependence of the fractal
dimension was linked to the dynamics of two cluster populations, small
clusters ($k \leq 15$) at early times and large clusters ($k > 15$) at late
times. The average cluster fractal dimension, thus, was related to the dominant
agglomeration mechanism. The generated aggregates had a cluster coordination
number, defined as the mean number of first neighbours of a monomer in a cluster,
of more than five (at late times) suggesting that the clusters were compact
and not porous. The generated clusters were long, compact, and tubular with
a high mean coordination number and low fractal dimension.

We argued that these two salient features of aggregate morphology,
small-scale local compactness (evident at the early stages
of the agglomeration process) and longer-scale tubular structure
(evident at later stages of the agglomeration process),
were a consequence of properties of the monomer-monomer interaction
potential.
The small-scale structure is determined by the isotropy of the
potential that allows short-time local reorientation of bonded
monomers induced by the thermal noise. Monomers are free to
rearrange to maximize their interaction with other monomers, since
there is no angular potential to hinder bending of monomer-monomer
bonds, under the constraint that monomer-monomer bonds are not
stretched. The large-scale structure
is determined by the potential interaction range.
Colliding, locally compact, clusters rearrange locally at the point of contact:
the attractive interaction range of the potential used in the simulations
was too short to induce larger-scale rearrangement. Increasing
the attractive range would lead to more spherical aggregates.
After the initial rearrangements the aggregates remained rigid till the
next collision.

The dynamic cluster properties were analyzed in terms of the cluster translational
diffusion coefficient. We found that the diffusion coefficient of a $k$-mer
scaled with cluster size as $D_k\propto k^{-1}$: aggregates diffuse
like massive monomers. Furthermore, the average (per-unit-mass) friction
coefficient of a monomer in a $k$-monomer cluster was found to equal the
(per-unit-mass) friction coefficient of an isolated monomer. Hence, the friction coefficient
of a monomer in a cluster was determined to be independent of its state of aggregation,
an approximation referred to as free draining approximation:
the average monomer shielding factor of the generated clusters was unity. We argued
that this diffusive behaviour is a consequence of the
absence of shielding in the Langevin equations. This is an inevitable consequence
of the usual application of Langevin simulations, unless a (time dependent) shielding coefficient is
explicitly introduced in random force term of the Langevin equations.
The shielding coefficient would, as a consequence of the Fluctuation
Dissipation Theorem, modify the monomer friction coefficient. We, thus, referred to these
clusters as \textit{ideal clusters}, with respect to their transport properties.
Similarly, the generated clusters did not have, on average,
a preferred orientation.

We also calculated numerically and compared to
analytical predictions kernel elements $K_{ij}$ for low
indexes. An extensive numerical investigation of the kernel
elements  is beyond the purpose of the present study, but it can be performed
using the methodology described herein.

We conclude that aggregates generated by unshielded Langevin equations
of motion of monomers interacting via a central, rapidly decaying potential
are on small scales, locally compact, on larger scales tubular, and
they diffuse as massive monomers.

\clearpage

\begin{table}
\caption{Overview of cluster static properties.}
\smallskip

\begin{center}
\begin{tabular}{|c|c|c|c|} \hline \hline
& Fractal dimension & Prefactor ($R_g = a k^{1/d_f}$)
& Prefactor  ($k = k_g (2R_g)^{d_f}$) \\ \hline \hline
All clusters & $d_f = 1.62 \pm 0.02$ & $a = 0.242 \pm 0.006$ & $k_g = 3.24$ \\
$k \le 15$ & $d_f^{mc} = 2.25 \pm 0.05$ & $a_{mc} = 0.386 \pm 0.009$ & $k_g^{mc} = 1.75$ \\
$ 15 < k$ & $d_f^{cc} = 1.56 \pm 0.02$ & $a_{cc} = 0.218 \pm 0.007$ & $k_g^{cc} = 3.65$ \\
\hline \hline
\end{tabular}
\label{tab:RadiusScaling}
\end{center}
\end{table}

\bigskip

\begin{table}
\caption{Overview of cluster diffusive properties. The percentage in
parenthesis (fifth column) is the relative difference of the two
mobility radii reported in the fourth and fifth columns.}
\smallskip

\begin{center}
\begin{tabular}{|c|c|c|c|c|c|c|c|} \hline \hline
& & $k=4$ & $k=10$ & $k=50$ & $k=98$  \\ \hline \hline
$D_k$ & Simulations, Eq.~(\ref{eq:diffusion_simu_cluster}) & $1.30\cdot 10^{-1}$ & $4.92\cdot 10^{-2}$ & $9.48\cdot 10^{-3}$
& $4.84\cdot 10^{-3}$  \\ \hline
$R_g$ & Simulations, Fig.~(\ref{two_df}) & $7.5\cdot 10^{-1}$ & $1.14$ & $2.12$ & $4.70$  \\ \hline
$\beta_k$ & Simulations, Eq.~(\ref{eq:cluster-diffusion-coefficient})  & $9.6\cdot 10^{-1}$ & $1.01$ & $1.05$ & $1.05$  \\ \hline
$R_m$ & Simulations, Eq.~(\ref{eq:mobility_radius}) & $1.92$ & $5.08$ & $26.35$ & $51.56$ \\ \hline
$R_m$ & Eq.~(\ref{eq:DiffusionShielding}) with $\eta_k = 1$ &$2$ ($4\%$) & $5$ ($1.6\%$) & $25$ ($5.4\%$) & $49$ ($5.2\%$) \\
\hline \hline
\end{tabular}
\label{table_diffusion}
\end{center}
\end{table}
\bigskip

\begin{table}
\caption{Comparison of kernel elements (expressed in units of $8 \pi D_1$).}
\smallskip

\begin{center}
\begin{tabular}{|c|c|c|c|c|c|c|} \hline \hline
($i,j$) & {$K_{ij}^{LD}\; \rm  [Eq.\; (\ref{eq:kernel_complete})]$} & {$K_{ij}\; \rm
[Eq. \; (\ref{eq:kernel_nonlin_aver})]$}   \\ \hline \hline
(${1,1}$) & $0.213$ & $0.214$   \\ \hline
(${1,2}$) & $0.294$ & $0.298$  \\ \hline
(${1,3}$) & $0.309$ & $0.397$   \\ \hline
(${1,4}$) & $0.309$ & $0.449$  \\ \hline
(${2,2}$) & $0.317$ & $0.317$  \\ \hline
(${2,3}$) & $0.303$ & $0.349$  \\ \hline
\hline
\end{tabular}
\label{table_kernel}
\end{center}
\end{table}


\clearpage

\clearpage

\clearpage

\clearpage

\clearpage

\clearpage

\clearpage

\clearpage

\clearpage

\end{document}